\documentclass[11pt]{article}
\usepackage[utf8]{inputenc}
\usepackage{graphicx}
\usepackage{geometry}
\usepackage{amsmath,amssymb}

\usepackage{comment}

\DeclareMathOperator{\logit}{logit}
\DeclareMathOperator{\expit}{expit}

\DeclareMathOperator*{\argmax}{arg\,max}

\usepackage[backend=biber,style=nature, maxbibnames=990]{biblatex}
\title{\Huge Collective inference of the truth of propositions\\from crowd probability judgments}

\author{Patrick Stinson, Jasper van den Bosch, Trenton Jerde, Nikolaus Kriegeskorte}

\date{January 2025}

\begin{document}

\maketitle

\begin{abstract}
    \noindent Every day, we judge the probability of propositions. When we communicate graded confidence (e.g. ``I am 90\% sure"), we enable others to gauge how much weight to attach to our judgment. Ideally, people should share their judgments to reach more accurate conclusions collectively. Peer-to-peer tools for collective inference could help debunk disinformation and amplify reliable information on social networks, improving democratic discourse. However, individuals fall short of the ideal of well-calibrated probability judgments, and group dynamics can amplify errors and polarize opinions. Here, we connect insights from cognitive science, structured expert judgment, and crowdsourcing to infer the truth of propositions from human probability judgments. In an online experiment, 376 participants judged the probability of each of 1,200 general-knowledge claims for which we have ground truth (451,200 ratings). Aggregating binary judgments by majority vote already exhibits the ``wisdom of the crowd''---the superior accuracy of collective inferences relative to individual inferences. However, using continuous probability ratings and accounting for individual accuracy and calibration significantly improves collective inferences.
    Peer judgment behavior can be modeled probabilistically, and individual parameters capturing each peer's accuracy and miscalibration can be inferred jointly with the claim probabilities. This unsupervised approach can be complemented by supervised methods relying on truth labels to learn models that achieve well-calibrated collective inference. The algorithms we introduce can empower groups of collaborators and online communities to pool their distributed intelligence and jointly judge the probability of propositions with a well-calibrated sense of uncertainty.

\end{abstract}

\section*{Introduction}
Judgments of probability are necessary to make accurate inferences and sound decisions. Each day, we make many of these judgments and base our decisions on them. People express their judgments using a nuanced range of categories of certainty (e.g. ``certainly", ``probably", ``I'm pretty sure...") or using numerical percentages (``I'm 90\% sure...") \cite{wallstenComparingCalibrationCoherence1993,wallstenPreferencesReasonsCommunicating1993,wallstenStateArtEncoding1983}. Graded confidence is required for optimal decisions under uncertainty \cite{maBayesianDecisionModels2019}. Although people rely on heuristics and are affected by biases \cite{tverskyJudgmentUncertaintyHeuristics1974} and overconfidence \cite{harveyConfidenceJudgment1997}, in many domains, their judgments can be understood as approximations to the normative ideal of probabilistic inference \cite{simonModelsMan1957, simonModelsBoundedRationality1982, tenenbaumTheorybasedBayesianModels2006,griffithsProbabilisticModelsCognition2010, griffithsOptimalPredictionsEveryday2006, gershmanComputationalRationalityConverging2015}. 

When we communicate our graded confidence through language, we enable our peers to gauge how much weight to attach to our judgment. Combining the judgments of a group can yield a collective judgment that is more accurate than the individual judgments  \cite{galtonVoxPopuli1907}, a phenomenon known as the ``wisdom of the crowd" \cite{surowieckiWisdomCrowds2004}.  In practice, however, social processes that lead to collective judgments often go astray \cite{kerrGroupPerformanceDecision2004,almaatouqAdaptiveSocialNetworks2020,bailExposureOpposingViews2018,pennycookPsychologyFakeNews2021}.

These issues gain urgency in the context of modern web technology, which connects us but lacks mechanisms that would enable an online community to collectively achieve accurate judgments of the probability of questionable claims \cite{lorenz-spreenHowBehaviouralSciences2020,lazerScienceFakeNews2018,stewartInformationGerrymanderingUndemocratic2019,delvicarioSpreadingMisinformationOnline2016,schmidtAnatomyNewsConsumption2017, vosoughiSpreadTrueFalse2018, shiWisdomPolarizedCrowds2019, kozyrevaCitizensInternetConfronting2020,geersOnlineMisinformationEngagement2024,,pennycookFightingMisinformationSocial2019}. Social media enable each of us to share and broadcast emotional responses to online information with a single click, such as a ``like". A similarly efficient mechanism for sharing cognitive responses, such as probability ratings, might help a community debunk false information, amplify accurate information, and engage the continuum between these extremes with a well-calibrated sense of uncertainty.

Here, we investigate the idea that a group of people, such as a panel of experts, a group of researchers, or an online community of citizens, can collectively evaluate a set of claims by annotating them with probability ratings. We refer to the group members as ``peers" to emphasize the equal status of the contributors. 
We compare existing and novel algorithms that provide collective inferences by combining human judgments. 

How to combine probability judgments has been explored in the fields of probabilistic opinion pooling \cite{degrootReachingConsensus1974,dietrichProbabilisticOpinionPooling2017,genestCombiningProbabilityDistributions1986}, structured expert judgment \cite{haneaExpertJudgementRisk2021,ohaganExpertKnowledgeElicitation2019}, and forecasting \cite{ungarGoodJudgmentProject2012, turnerForecastAggregationRecalibration2014,clemenCombiningForecastsReview1989}. These fields have developed methods for the elicitation and aggregation of probability judgments. Our approach builds on those literatures and connects them to crowdsourcing and collaborative filtering \cite{aggarwalRecommenderSystemsTextbook2016, zhengTruthInferenceCrowdsourcing2017}, where large volumes of judgments are modeled probabilistically as resulting from the interaction between properties of the people making the judgments and the items judged. Our approach also has connections to variants of item response theory that involve inferences about items as well as participants \cite{hambletonItemResponseTheory1985, embretsonItemResponseTheory2013}. 

Collective inference of the probability of claims from a claim-by-peer matrix of human probability ratings is a fundamental problem with transformative potential for social media and other applications where large volumes of claims are to be evaluated collaboratively by groups of people. As in opinion pooling and structured expert judgment, this form of collective inference requires combining  probability judgments and can benefit from a supra-Bayesian approach \cite{keeneyDecisionsMultipleObjectives1976,winklerConsensusSubjectiveProbability1968}, where the human judgments form the data. As in crowdsourcing \cite{smythInferringGroundTruth1994,raykarLearningCrowds2010} and collaborative filtering \cite{resnickGroupLensOpenArchitecture1994, sarwarItembasedCollaborativeFiltering2001}, the methods must work for large, sparse matrices of non-expert judgments, and inference should ideally be based on a model of how items and people (claims and peers here) interact to generate the data (judgments).

\begin{figure}[b!]
    \centering
    \includegraphics[width=0.8\textwidth]{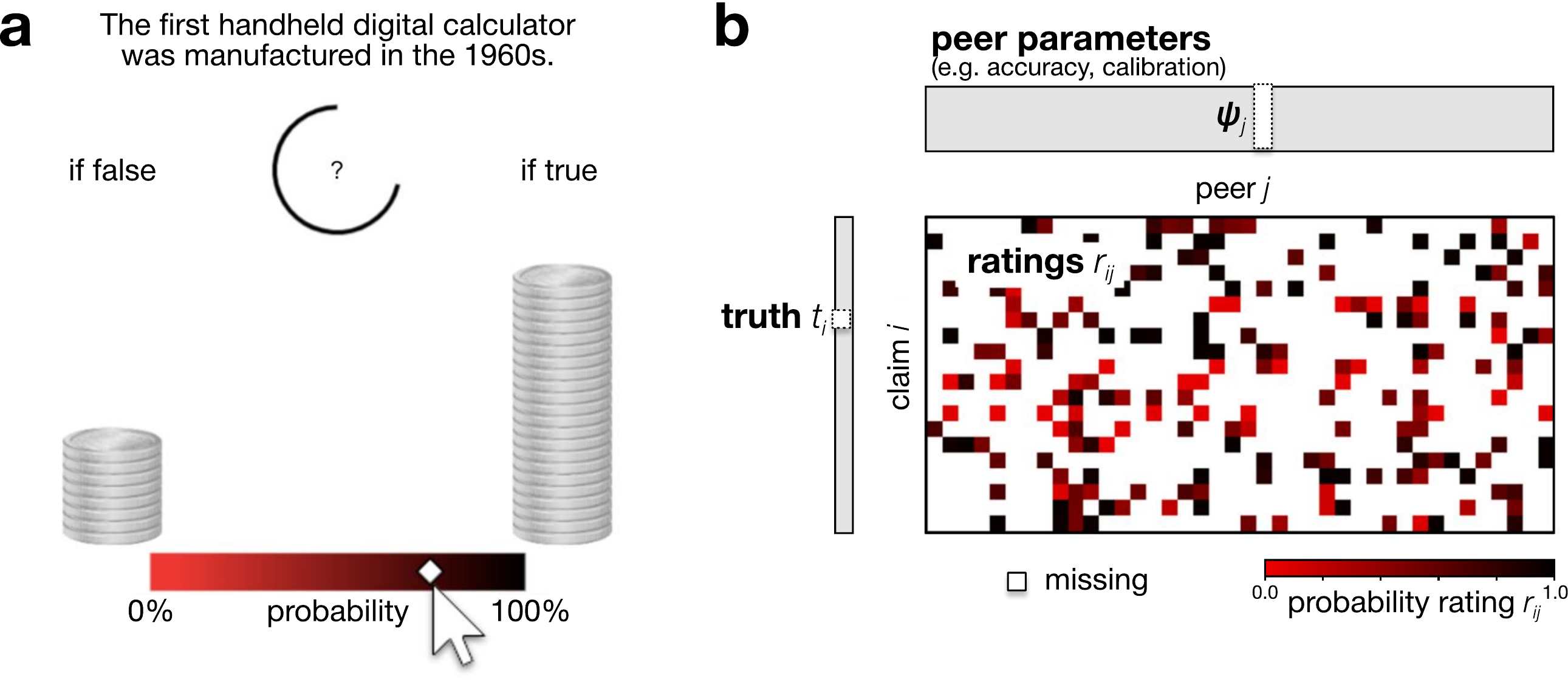}
    \caption{\footnotesize \textbf{Human probability judgments and collective-inference problem.} (\textbf{a}) Judgment acquisition in the online experiment. A screen view for an example trial as experienced by participants is shown. Participants were instructed to click on the probability scale bar within a time limit of 20 s (visualized as a diminishing circular arc around the question mark) to record their rating. The coin stacks to the left and right represented rewards for submitting a rating at the current mouse position. The participant received the left stack if the claim was false and the right stack if the claim was true. The coin stacks changed size dynamically with horizontal mouse position when hovering over the rating bar, such that the stack closer to the mouse pointer is larger (i.e. larger reward received when the rating is on the correct side of 50\%). For the full range of feedback and reward conditions (including reward functions for which well-calibrated ratings, underconfident ratings, or overconfident ratings maximize the expected reward), see \textit{Methods}. Coin rewards were imaginary and not reflected in the payments to participants. Each participant was randomly assigned to one of the feedback conditions and rated all 1,200 general-knowledge claims in randomized order over the course of six sessions. (\textbf{b}) Probability ratings matrix. The collective-inference problem is to infer the probability $p(t_i)$ of each claim given a (possibly sparse) matrix of probability ratings $r_{ij}$, where $t_i$ is the binary truth value of claim $i$ and $r_{ij}$ is the rating peer $j$ has given for claim $i$. The claim probabilities can be inferred by learning peer parameters $\psi_j$, reflecting the judgment behavior of each individual peer (e.g. accuracy and calibration). The matrix shown is sparse (many missing values, white) to illustrate the inferential challenge faced in a real-world application. In the experiment, a dense matrix was acquired (each of the 376 participants rated each of the 1,200 claims). To assess the ability of collective-inference algorithms for sparse matrices, we resample the data to simulate inference challenges realistic for applications.}
    \label{fig:fig1}
\end{figure}

We implemented (1) previously described simple heuristic aggregation methods, (2) a novel supra-Bayesian inference algorithm that learns a probabilistic generative model of each peer's judgment behavior to infer the probability of each claim without requiring any truth labels, and (3) supervised models that rely on truth labels for a training set of claims to infer the probability of new claims from the human ratings. These collective-inference algorithms could be scaled to large numbers of people, e.g. on social media. We evaluate and compare these algorithms using human probability ratings of general knowledge claims acquired in an online experiment. Each of 376 online participants was presented with each of 1,200 claims and responded within a time window of 20 seconds by clicking on a probability bar ranging from 0 to 1, yielding a total of 451,200 ratings (Fig. \ref{fig:fig1}). Because we know the truth value of each claim, the data set enables us to objectively evaluate and compare the collective-inference algorithms. 

This study makes the following contributions: (1) We introduce an algorithm that combines judgment-generative models and inference by expectation maximization (EM) to jointly infer individual peer behavior and the probability of the claims. (2) We provide a data set much larger than those previously available of 451,200 human probability judgments,  including 376 raters each of whom judged each of 1,200 claims (half of them true, the other half false) across six sessions, enabling detailed modeling of individual rating behavior. (3) We perform the first crossvalidated inferential comparison of a wide range of algorithms for combining human probability judgments, treating both claims and peers as random effects. The inferential comparisons use a two-factor (claim by peer) bootstrap procedure that we recently introduced \cite{schuttStatisticalInferenceRepresentational2023}, taking advantage of the fully crossed new data set, and reveal how different methods perform when given different numbers of ratings as input. (4) We demonstrate that collective inference benefits from the modeling of individual accuracy and miscalibration and how this can be achieved with or without truth labels. We also demonstrate the benefits of continuous (over binary) judgments and inferentially compare a wide range of simpler judgment aggregation methods.

The collective inference methods developed here will be useful for groups of people across scales, ranging from panels of experts to labs, teams of analysts, institutes, companies, news organizations, and online communities on social media \cite{wheatleyEmergingScienceInteracting2023, bailBreakingSocialMedia2021}. The human data, collective-inference algorithms, and statistical inference methods for comparing algorithms will be openly accessible.

\section*{Results}
\subsection*{Majority vote demonstrates the wisdom of the crowd}
We first consider the accuracy achieved by simple methods for aggregating the judgments. The accuracy of an individual person or collective inference algorithm is defined as the rate of correct answers. An estimated probability is counted as correct if it falls on the right side of 0.5. If we chose a single peer's probability rating at random, the accuracy of our collective inferences about the claims would match the average accuracy of the peers: about 62\% for the set of 1,200 general-knowledge claims in our online experiment (Fig. \ref{fig:fig2}). Trusting a random peer does not benefit from the wisdom of the crowd.

A simple method to aggregate multiple ratings is the majority vote. We first binarize the probability ratings by thresholding them at 0.5, so as to determine whether the peer considered the claim to be more likely to be true or more likely to be false. We consider the claim true if the number of ratings greater than 0.5 exceeds the number smaller than 0.5. We consider the claim false if ratings below 0.5 dominate.  (Ratings of exactly 0.5 are not counted, and in case the numbers of votes for and against the claim are equal, we perform a random tie break.) The accuracy of the majority vote approaches 70\% when 10 or more ratings per claim are used (Fig. \ref{fig:fig2}). The majority vote is significantly more accurate than trusting a random peer's rating when $3$ or more ratings per claim are available (Fig. \ref{fig:fig2}, paired one-tailed $t$-test, $p<0.05$, Bonferroni-corrected for $8$ different numbers of ratings per claim). This provides a first simple illustration of the wisdom of the crowd \cite{surowieckiWisdomCrowds2004,turnerWisdomCrowdApproach2011}.

All statistical comparisons of collective-inference algorithms in this paper rely on a 2-factor bootstrap procedure that treats both peers and claims as random effects. See \textit{Methods} for statistical procedures and \textit{Supplementary Information} (section \textit{Idiosyncrasies of random ratings, majority vote, and median rating}) for discussion of the case of two ratings per claim and of the median rating.

\subsection*{Averaging of continuous ratings beats counting of binary votes}
\begin{figure}[h!]
    \centering
    \includegraphics[width=\textwidth]{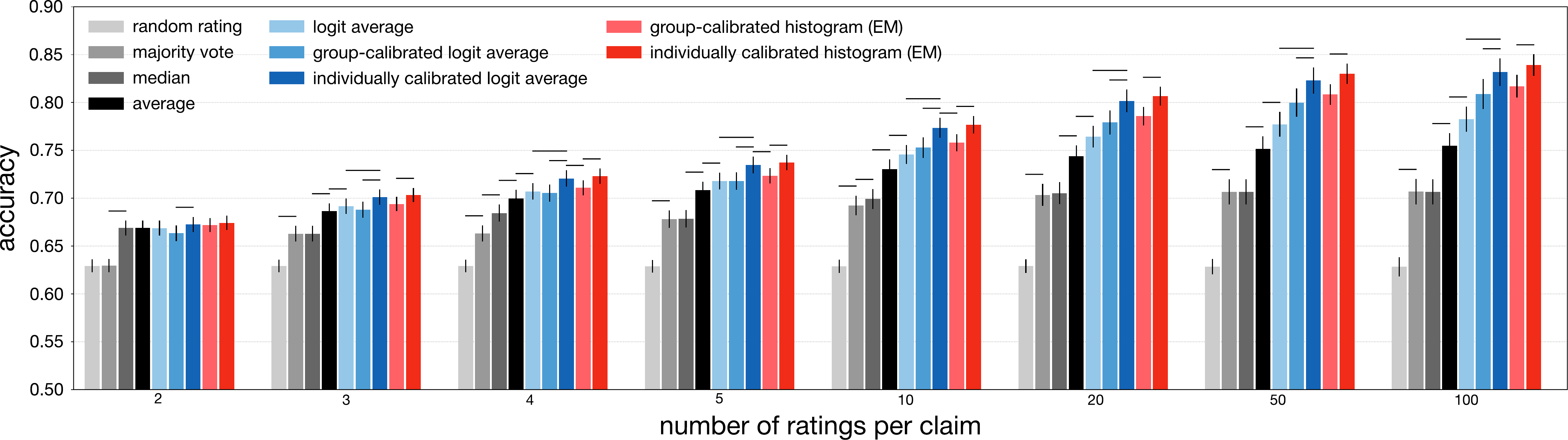}
    \caption{\footnotesize \textbf{Accuracy of collective-inference algorithms for different numbers of ratings per claim.} Bars show the accuracy of a progression of collective-inference algorithms (colors). For neighboring bars, horizontal lines show significant differences (2-factor bootstrap, generalizing across both peers and claims, $p<0.05$, Bonferroni-corrected for 8 comparisons across numbers of ratings per claim, one-sided test for each pair of adjacent models of the hypothesis that the more sophisticated model is better). The majority vote (gray) outperforms a random peer (light gray) for $>2$ ratings per claim. The rating average (black) outperforms the majority vote for $>4$ ratings per claim. The logit average (light blue) outperforms the rating average for $>2$ ratings per claim. Calibrating the ratings at the group-level (mid-blue) does not improve performance. Calibrating the ratings at the individual level (dark blue) improves performance for 3 to 50 ratings per claim. The calibrated logit methods require a separate set of truth-labeled claims to estimate the calibration function for each peer. Individual inaccuracy and miscalibration can also be accounted for without truth labels, by inferring the probability of the claims using individual judgment-generative models fitted using expectation maximization (red), which achieves similar performance.
    Error bars represent standard error of the mean accuracy under 2-factor bootstrap resampling, reflecting measurement error as well as random sampling of both peers and claims.}
\label{fig:fig2}
\end{figure}

The majority vote binarizes the ratings, which removes information. Extreme ratings closer to 0 or 1 reflect greater confidence than ratings close to 0.5. A simple aggregation rule that gives greater pull to extreme ratings is the rating average. If we average $10$ ratings per claim, the accuracy of our collective inferences increases to about 73\%, and averaging $100$ ratings per claim yields about 75\% accuracy. The rating average is significantly more accurate than the majority vote when $3$ or more ratings per claim are used (Fig. \ref{fig:fig2}). This provides a first indication that the information about confidence contained in continuous probability ratings is useful for collective inference. (The advantage of using continuous rating information is also evident in the context of inference using judgment-generative models and discriminative supervised models. These results are shown in Fig. \ref{fig:fig4} and are described below.)

Averaging might be a good approach if each peer judged on the basis of the same evidence. In particular, if each peer used a sampling algorithm for computing a posterior probability for each claim and computed the same number of samples, then the average of the probability ratings would give the posterior for the pool of all samples computed in a distributed fashion by the crowd as a whole.

\subsection*{Independent opinion pool beats rating average}
If the probability ratings do not all reflect the same evidence, then probability averaging is not the optimal way of combining the ratings. Let us consider the opposite scenario where the probability ratings reflect independent evidence, and also assume, for the moment, that the ratings are well-calibrated. A well-calibrated rating is one that accurately reflects the peer's uncertainty, such that among all claims receiving rating $r$, the rate of true claims is $r$ (so, for example, among claims rated $r=0.8$, 80\% are true). If the peers' ratings are independent given the truth value of a claim and well-calibrated, the optimal aggregation rule is the Independent Opinion Pool \cite{bergerStatisticalDecisionTheory1985}: We multiply the binary probability distributions $[r_i, 1-r_i]$ across peers indexed by $i$ and renormalize the resulting pair of values to sum to unity (to provide a proper binary probability distribution). Equivalently, we can convert the probability ratings to logits (log odds), sum these, and convert back to a probability as our collective inference (details in \textit{Methods}). In terms of accuracy, the logit sum (i.e. the independent opinion pool) is equivalent to the logit average. In either case, the sign of the aggregate determines whether the inferred probability is greater than or less than 0.5. The logit average reaches an accuracy of about 77\% for 50 or more ratings per claim and is significantly more accurate than the rating average for $3$ or more ratings per claim (Fig. \ref{fig:fig2}). The effectiveness of the independent opinion pool suggests that there is some value in taking the ratings seriously as approximately calibrated indications of probability that are not entirely based on the same evidence.

Summing or averaging the logits yields collective inferences that are equivalent in terms of accuracy. However, if we sum the logits (implementing the independent opinion pool), the collective inferences are highly overconfident: For many ratings, the collective probability will be close to 0 or 1, and will not be well-calibrated. This indicates that the independent opinion pool's assumption of independent ratings (given the truth of the claim) is incorrect. If all peers instead drew from identical evidence, their ratings would covary given the truth value of the claim. Aggregation of the ratings might still be useful, but only to reduce any noise affecting the ratings. Rating noise might arise at the cognitive and/or motor level, causing random variation of the ratings. The assumption of noisy ratings that reflect the same evidence motivates using the average rather than the sum of the logits. We find that averaging the logits, instead of summing them, makes the collective inference underconfident (Supplementary Fig. \ref{fig:fig13}).

In reality, the evidence on which two people's ratings of the same claim are based is not expected to be identical or completely independent. Instead, we expect some unknown degree of overlap in the evidence people draw from. For a given claim, there is a limited pool of evidence. Each peer accesses some subset of the relevant facts, and the dependency between peer ratings may reflect the size of the evidence pool and the sources of information available to different peers.

A simple way to account for dependence is to use a convex combination of the logit sum and the logit average, dividing the sum, not by the actual number of ratings, but by an estimate of the effective number of independent ratings. We will return to this issue below in the section \textit{Calibrated collective inference requires a reference set of truth-labeled claims}.

\subsection*{Accounting for individual accuracy and calibration improves collective inference}
The independent opinion pool assumes that individual ratings are not only independent, but also well-calibrated. Human probability ratings are known not to be well-calibrated \cite{mellersIdentifyingCultivatingSuperforecasters2015,albrechtSimilarityupdatingModelProbability2021,pickhardtStudyPerformanceSubjective1974,stroopJudgmentGroupBetter1932}. Consistent with previous findings  \cite{harveyConfidenceJudgment1997,sollOverconfidenceIntervalEstimates2004}, our participants were overconfident on average.
An ideal collective-inference algorithm should correct individual biases in favor of high or low ratings, downweight inaccurate peers, and calibrate overconfident and underconfident peers, so as to optimally combine the ratings.



Bias, inaccuracy, and over- or underconfidence of individual peer judgments can be accounted for by estimating each peer's calibration function and using the estimate to calibrate the ratings before combining them. Instead of trusting a 0.8 rating to indicate a 0.8 probability of the claim, we can estimate how frequently claims rated around 0.8 by a particular peer are true. For each peer $j$, we need to model the calibration function, which specifies the probability 
$p_j(t=True|r_j)$ that a claim is true given that peer $j$ has given it rating $r_j$. Estimating a peer's calibration function requires that we have some information about the truth of the claims the peer has rated. If peer $j$ has rated a sufficient number of claims $i$ that we have truth labels $t_i \in \{True, False\}$ for, then we can estimate the calibration function. We can then sum or average the logits corresponding to $p_j(t=True|r_j)$ (for the different peers $j$ that have rated a claim) instead of the logits of the original ratings $r_j$.

Calibrating the individual peers entails that less informative peers have less influence. A peer whose ratings are unrelated to the truth of the claims will have calibrated logits equal to 0 and thus will not pull collective inference in either direction. More generally, a peer's calibrated logits will accurately reflect her actual uncertainty. 

To estimate the effect of calibrating the logits, we used separate training and test sets of claims for the same set of 376 peers. We designated a random subset of 600 claims as the training set, using these claims' truth labels to estimate the logistic calibration function $p_j(t=True|r)=\expit[\logit[r]/c_j-b_j]$ for each peer $j$, where $b_j$ is the peer's bias and $c_j$ is the peer's confidence (the factor by which the peer inflates the evidence). We used the ratings by these peers of the other 600 claims as the test set to estimate the accuracy of the individually calibrated logit average (blue in Fig. \ref{fig:fig2}). To simulate performance for different numbers of ratings per claim, we sparsified the data (see \textit{Methods}). The relatively large number (600) of labeled training claims was chosen to provide an estimate of the potential of individual calibration under ideal conditions.

Averaging individually calibrated logits yielded an accuracy of about 82\% when 50 or more ratings per claim were used. The individually calibrated logit average outperformed uncalibrated logit average, with the difference significant whenever 3 or more ratings per claim were used for collective inference ($p<0.05$ for 3, 4, 5, 10, 20, 50, and 100 ratings per claim, Bonferroni-corrected for 8 different numbers of ratings per claim; Fig. \ref{fig:fig2}). 

Calibration only improves collective inference if it accounts for individual differences among peers. When we calibrated each peer's ratings using the calibration parameters estimated for the group as a whole (mid-blue bars in Fig. \ref{fig:fig2}), calibration did not yield an advantage over the uncalibrated logit average. The individually calibrated logit average yielded higher accuracy than the group-calibrated logit average for all tested numbers of ratings per claim ($p<0.05$ for all tested numbers of ratings per claim, Bonferroni-corrected for 8 different numbers of ratings per claim; Fig. \ref{fig:fig2}). These results indicate that collective inference should account for individual differences in judgment behavior.

\subsection*{Judgment-generative models enable collective inference without truth labels}
The individually calibrated logit average requires that each peer has rated a substantial number of claims for which we have truth labels. In many applications, we will not have truth labels at all or not for the claims a particular peer has rated. Ideally, we would like to be able to solve the chicken-and-egg problem of inferring the probabilities of the claims and the propensities of the peers jointly. 


The normative approach is to learn a generative model $p_{\psi_j}(r|t)$ specifying each peer $j$'s probability density over ratings given the binary truth of a claim. The parameter vector $\psi_j$ captures the rating behavior of peer $j$. We find that using a histogram to represent each peer's truth-conditional rating distribution works well in practice (see \textit{Methods} for details). We use the expectation-maximization (EM) algorithm to alternately infer the probabilities of the claims and the parameters $\psi_j$ capturing each peer $j$'s rating behavior given a true or false claim. 

\begin{center}
    \begin{figure}[h!]
        \centering
        \includegraphics[width=\textwidth]{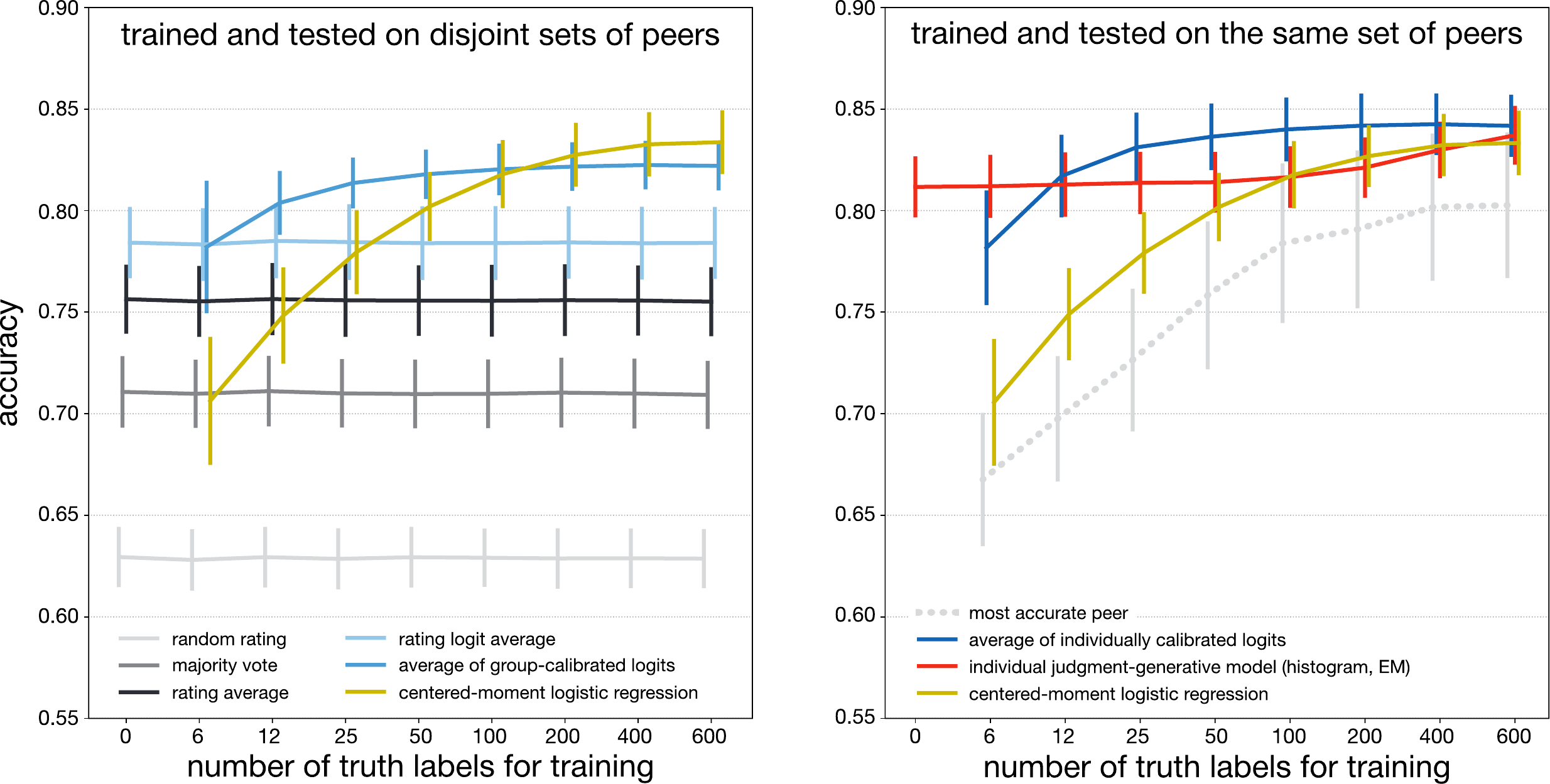}
        \caption{\footnotesize\textbf{Inferring truth without peer-specific training.} Given a training set of ratings and truth labels of the rated claims, we may want to infer the truth of new claims from ratings by new peers. In this scenario, algorithms must generalize simultaneously to new claims and new peers, and since we may have only 1 rating from each of the new peers, algorithms cannot use peer-specific parameters. The left panel shows the accuracy (vertical axis) of inference algorithms that can be applied to ratings from new peers each of whom has only provided a single rating. The naive methods (gray to black) and the rating logit average (light blue) do not require any training and so their accuracy does not depend on the size of the training set of labeled claims (horizontal axis). The average of group-calibrated logits requires training data to learn a group-level logistic calibration model. It outperforms the previous methods for training sets as small as a few dozen truth-labeled claims. When 100 or more truth-labeled claims are available, centered-moment logistic regression (green) becomes competitive and for hundreds of truth-labeled claims may slightly outperform the average of group-calibrated logits. The right panel shows performance of centered-moment logistic regression (green) in the context where truth-labeled claims are available for the same peers whose ratings our inferences are based on. In this scenario, we can learn individual peer parameters with models such as average of individually calibrated logits (dark blue) and individual judgment-generative model (histogram, EM; red). These peer-specific algorithms dominate unless 100 or more labeled claims are available. For both panels, the training and test sets each consist of 188 peers and 600 claims. The number of ratings per claim in these analyses is 188. See Supplementary Fig. \ref{fig:fig14} for the performance of centered-moment logistic regression on a range of numbers of ratings per claim. For the right panel, training and test sets contain ratings from the same 188 peers, but of different claims. For the left panel, training and test sets contain ratings of different claims by different peers. The horizontal axis shows the number of truth-labeled claims randomly chosen from the training set. The test accuracy of trained inference models is shown for the scenario in which each peer has rated each claim. Plots for different algorithms are slightly shifted horizontally to show the error bars, which represent the standard error of the mean, estimated by 2-factor bootstrap resampling of claims and peers.}
            \label{fig:fig3}
    \end{figure}
\end{center}

The judgment-generative model achieves an accuracy of about 83\% without using any truth labels when 50 or more ratings per claim are used. The judgment-generative model (histogram EM, dark red in Fig. \ref{fig:fig2}) matches the individually calibrated logit average (dark blue in Fig. \ref{fig:fig2}), despite using no truth labels. The inferential comparison revealed no significant differences in accuracy for any of the 8 numbers of ratings per claim (Supplementary Fig. \ref{supp_fig:pairwise_comparison}). Joint inference of claim probabilities and peer propensities, thus, is a highly attractive approach for collective inference. If we have truth labels for some of the claims, the corresponding probabilities can be set to 0 or 1 in the EM inference, which can further improve collective inference (red line in Fig. \ref{fig:fig3}, right panel).

We saw above that averaging the ratings or their logits outperformed counting binarized ratings (majority vote). An important question is whether continuous ratings also yield collective inferences superior to those based on binary responses when using judgment-generative probabilistic models. Our judgment-generative model predicts a peer-specific truth-conditional probability density over continuous ratings. An influential judgment-generative model that predicts peer-specific truth-conditional probabilities of binary responses has been proposed by Dawid and Skene 
\cite{dawidMaximumLikelihoodEstimation1979}. Our continuous-response model significantly outperforms this binary response model (Fig. \ref{fig:fig4}, probabilities binarized by thresholding at $0.5$). This finding provides further support for the hypothesis that continuous probability ratings provide better information for collective inference than binary responses.




\subsection*{Supervised logistic regression enables collective inference with 1 rating per peer}
Inferring claim probabilities and individual behavior jointly with a judgment-generative model for each peer does not require truth labels, but it does require a sufficient number of ratings from each individual. What if we have only a single rating from each peer? In that case, it is not possible to learn a model of each peer's rating behavior. We could use the logit average in this scenario. However, we might be able to do better than the logit average by supervised learning (using truth labels) of a mapping from a set of ratings to the claim probability. 

If we have a training set of ratings of truth-labeled claims, we can use supervised machine learning to predict the probability of a claim from a set of ratings without any modeling of individual peer behavior. One approach that works well is logistic regression on the basis of the centered moments of the ratings. We first compute the mean of the ratings, then center the ratings on this mean. We then compute the mean square (variance), the mean cube (skewness), the mean 4-th power (kurtosis), and the mean 5-th power of the centered ratings. These five numbers characterize the location and shape of the ratings distribution for a claim. A linear logistic regression model takes the five moments as input and assigns a probability to the claim.

When trained with a representative set of peers and claims, centered-moment logistic regression can perform surprisingly well. Trained with a data set of 100 or more ratings of truth-labeled claims, the model performs competitively when given enough ratings per claim for collective inference. Fig. \ref{fig:fig3} shows how the accuracy of centered-moment logistic regression improves as the training set of truth-labeled claims grows, relative to the other algorithms. Collective inference in these results relies on 188 ratings per claim and performance plateaus at about 83\% accuracy. For 75 ratings per claim for collective inference, centered-moment logistic still achieved an average accuracy of about 83\% (Supplementary Fig. \ref{fig:fig14}, which shows the dependence on the number of ratings per claim used for inference),  significantly higher than the group-calibrated judgment-generative model, the group-calibrated logit-average, and the uncalibrated logit-average ($p < 0.05$, 100 bootstrap resamplings of both peers and claims and a randomized train/test split consisting of 100 training claims with known truth value and 1,100 test claims). However, when given 25 ratings per claim or less for collective inference, centered-moment logistic regression was no longer significantly more accurate. Supplementary Fig. \ref{fig:fig14} shows comprehensive performance results (accuracy, area under the receiver-operating characteristic, and Brier score) for different numbers of truth-labeled training examples (one from each peer of a separate training set of peers) and different numbers of ratings per claim for collective inference. These results show that a sufficiently large set of ratings (with truth labels for at least 100 of the claims) can be useful for collective inference even if we only have one rating per peer and therefore cannot leverage peer-specific models.

\begin{center}
    \begin{figure}[h!]
        \includegraphics[width=1.0\textwidth]{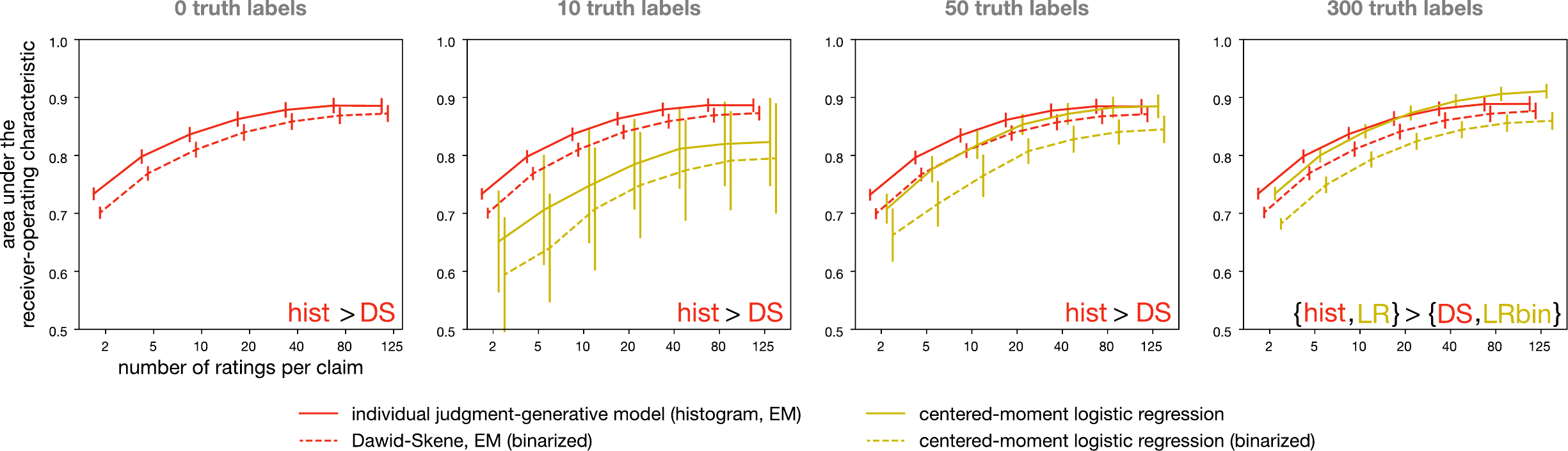}\
        \caption{\footnotesize\textbf{Probability ratings are superior to binary truth judgments for collective inference with unsupervised generative and supervised discriminative models.} Performance of supervised and unsupervised models for continuous (solid lines) and binarized ratings (dashed lines), as a function of the number of truth labels constraining the models (left panel: 0 truth labels, right panel: 300 truth labels) and the number of ratings per claim (horizontal axes). Different numbers of ratings per claim were simulated by matrix resampling (Methods). The fewer ratings we have per claim, the sparser the ratings matrix. Models are learned using a dataset of 188 peers and 600 claims. The number claims provided with ground-truth labels differs across panels. Performance is reported for collective inference on another portion of the dataset whose labels the algorithms did not have access to. Plots are horizontally staggered to reveal the error bars, which represent the standard error. Significantly higher performance of model A than model B is indicated by A$>$B in each panel (pooled across all ratings per claim, paired \textit{t}-test, $p<.05$). Performance is measured as the area under the receiver-operating characteristic, where 0.5 is chance performance and 1.0 is perfect separation of true and false claims. For the corresponding analyses of accuracy and Brier score, see Supplmentary Fig. \ref{fig:continuousVsBinarized_accBrier}.}
        \label{fig:fig4}
    \end{figure}
\end{center}

Under a range of training set sizes (0, 10, 50, and 300 truth labels), centered-moment logistic regression performs comparably to the judgment-generative model, with the difference in performance not significant (Fig. \ref{fig:fig4}, $p>0.05$ 2-factor-bootstrap paired \textit{t}-test, df=187). A variant of the logistic regression model using binarized ratings performs significantly worse than the logistic regression model using continuous ratings with a training set of 300 claims (Fig. \ref{fig:fig4}, $p<.05$ 2-factor-bootstrap paired \textit{t}-test, df=187). This demonstrates the value of continuous probability ratings in the context of supervised models.

\subsection*{Calibrated collective inference requires a reference set of truth-labeled claims} 

The accuracy of collective inferences, which we have focused on thus far, provides one important indicator of collective-inference performance. It can be evaluated for algorithms that produce binary decisions, such as the majority vote, as well as for continuous probability estimates, where it is defined as the rate with which collective inferences fall on the correct side of 0.5. Probabilistic collective inferences, however, should be not only accurate, but also well-calibrated. 

\begin{center}
    \begin{figure}[h!]
        \centering
        \includegraphics[width=0.8\textwidth]{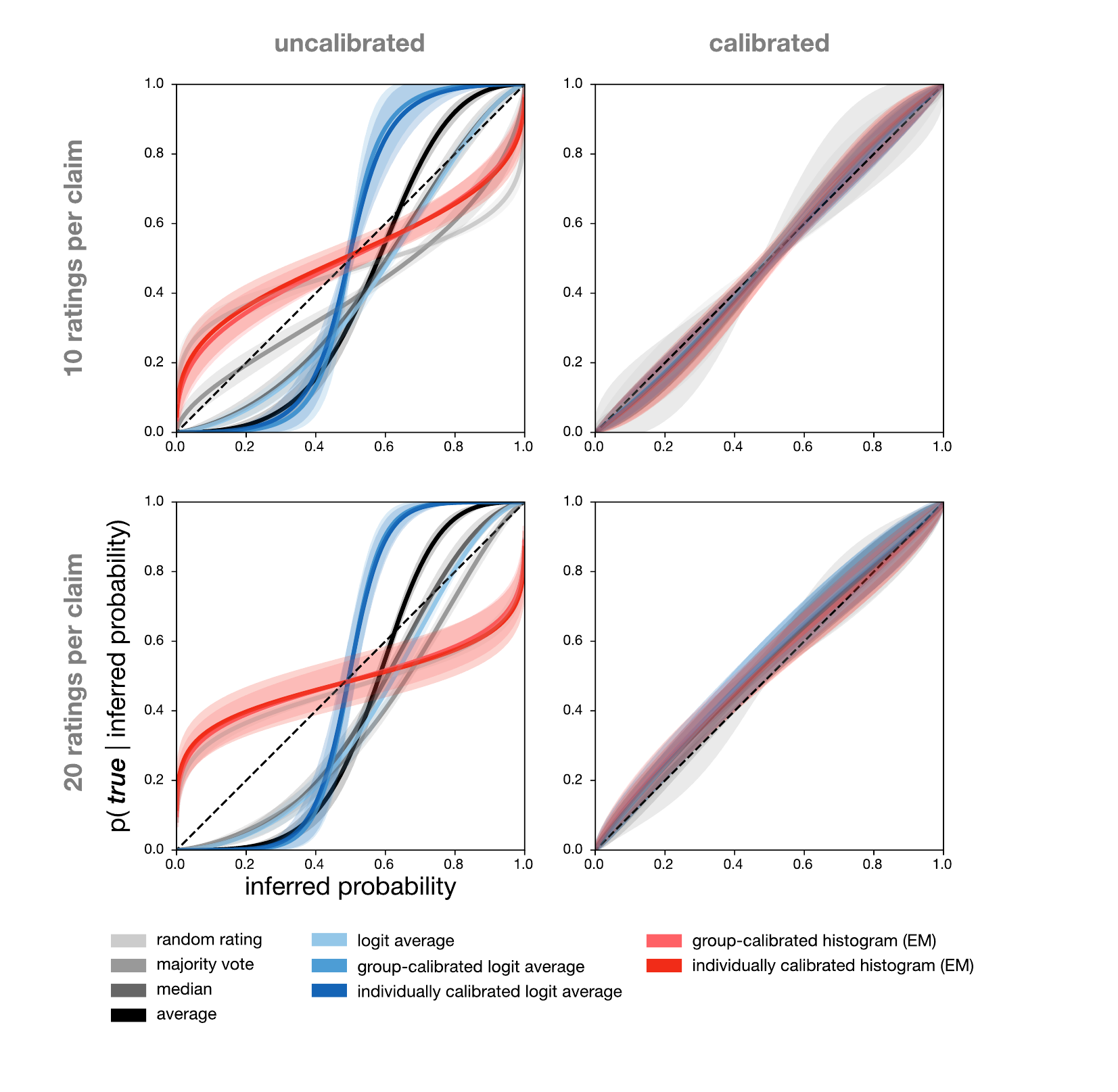}
        \vspace{-9mm}
        \caption{\footnotesize\textbf{Calibration of collective inference algorithms.} Collective inference algorithms using 10 and 20 peer ratings per claim (top row, bottom row) are trained on a random training set (100 labeled claims and 50 unlabeled claims). To assess calibration, we fitted a logistic model to the resulting inference estimates using a test set of 950 claims (thick lines). Deviations from the identity line reveal that collective inferences are substantially miscalibrated (left column). To remedy the miscalibration, we use a calibration set of 100 labeled claims (disjoint from both the training and the test set mentioned above). To calibrate each algorithm, we train a logistic regression model on the calibration set. The logistic regression model maps from uncalibrated to calibrated collective inferences. We then evaluate the calibration on the independent test set of 950 labeled claims. The right column shows that calibration using just 100 labeled claims generalizes to the test set. The evaluation procedure was performed using random splits of the total set of 1,200 claims, 100 times with thick lines representing the mean $+/-$ 1 standard deviation.}
        \label{fig:fig5}
    \end{figure}
\end{center}

We used the truth labels to fit the logistic calibration function (as already introduced above in the context of individual ratings) to the collective inferences $\hat{p}$ of different algorithms: $p(t=True|\hat{p}) = \expit[\logit[\hat{p}]/c-b]$. An algorithm is well-calibrated if the calibration function is close to the identity (with bias $b=0$ and confidence $c=1$). Results are shown in Fig. \ref{fig:fig5} as well as Supplementary Figures \ref{fig:fig12} and \ref{fig:fig13}. Choosing a random rating exhibits the peers' general overconfidence. The rating average is underconfident. As reported above, the logit sum (independent opinion pool) is overconfident, whereas the logit average is underconfident. The group- and individually calibrated logit averages are similarly underconfident, reflecting the assumption of entirely dependent ratings. The group and individual judgment-generative models are overconfident, because like the independent opinion pool they assume that ratings are conditionally independent given the truth of the claims.

A straightforward way to calibrate collective inferences is to pass the probability estimates through their calibration function (as we do when we calibrate the ratings of individual peers). This calibration step requires a set of truth-labeled claims. For collective inference algorithms that assume truth-conditional independence, calibration can correct for the overestimation of the evidence that results from the conditional dependency among ratings given the truth of the claims. We used a random subset of 50 training claims and 10 or 20 ratings per claim to calibrate the probability estimates of the algorithms. We then assessed the calibration on an independent test set of 950 different claims (Figure  \ref{fig:fig5}, Supplemental Figures \ref{fig:fig12} and \ref{fig:fig13}). Calibration generalized successfully to the test set: The test-set calibration curves closely track the identity line.

The centered-moment logistic regression model is trained with truth labels using the cross-entropy loss. This amounts to optimizing calibration on the training set. Centered-moment logistic regression, too, exhibited good calibration also on the test set (not shown).

\subsection*{Collective inference is robust to different motivations and perceived rewards} 

Humans vary in their probability judgment behavior \cite{leeInferringExpertiseKnowledge2012,albrechtSimilarityupdatingModelProbability2021,brennerOverconfidenceProbabilityFrequency1996,chaseVisionsRationality1998,genestCombiningProbabilityDistributions1986}. Beyond interindividual variation of traits and abilities, the behavioral context is known to affect judgments \cite{wallstenComparingCalibrationCoherence1993,klaymanOverconfidenceItDepends1999,hertwigDecisionsExperienceEffect2004,tsaiEffectsAmountInformation2008,hertwigFastFrugalHeuristics2009,muchnikSocialInfluenceBias2013}. 
A particular concern is that motivations created by feedback and by real or imagined rewards might distort ratings and create a challenge for collective inference. In order to assess the robustness of collective inference to variation in feedback and rewards, we randomly assigned each participant to one of seven feedback and reward conditions: In condition 1, no feedback was given. In condition 2, the correct answers were revealed after each block. In conditions 3-7, the correct answer was revealed after each trial. In condition 3, no other information was given. In conditions 4-7, subjects were additionally given the performance of other peers on the presented claim (condition 4) or imaginary coin rewards (conditions 5-7; Figure \ref{fig:fig1}a). The imaginary coin reward was chosen to encourage overconfident (condition 5), well-calibrated (proper scoring rule, condition 6), or underconfident (condition 7) ratings (details in \textit{Methods}, Table \ref{table:conditions}). These conditions were intended to enhance the variation of judgment behavior across participants and to enable more realistic and conservative estimates of the performance of collective-inference algorithms. Results presented thus far reflect the robustness of collective inferences to both interindividual variation and variation induced by the seven feedback and reward conditions. 

We statistically compared the performance of the logit average and the individually calibrated histogram algorithm (EM) across the seven feedback and reward conditions (Supplementary Figure \ref{fig:fig7}). We found no significant differences in collective-inference performance for any of the performance metrics (accuracy, auROC, Brier score; $p>0.05$, permutation $F$-test comparison of intra- and intergroup variance of performance metrics with 2-factor bootstrap, 10,000 draws). Algorithms accounting for individual peer behavior are expected to be robust not only to trait- and ability-related interindividual variation but also to context-dependent motivational variation (when the context is constant for each peer, as was the case here). For example, the influence of the different coin rewards on peer confidence would be counteracted by algorithms that correct individual miscalibration. However, even the logit average, which does not adapt to individual peer behavior, proved quite robust across feedback conditions. 

\begin{figure}
    \centering
    \includegraphics[width=.7\textwidth]{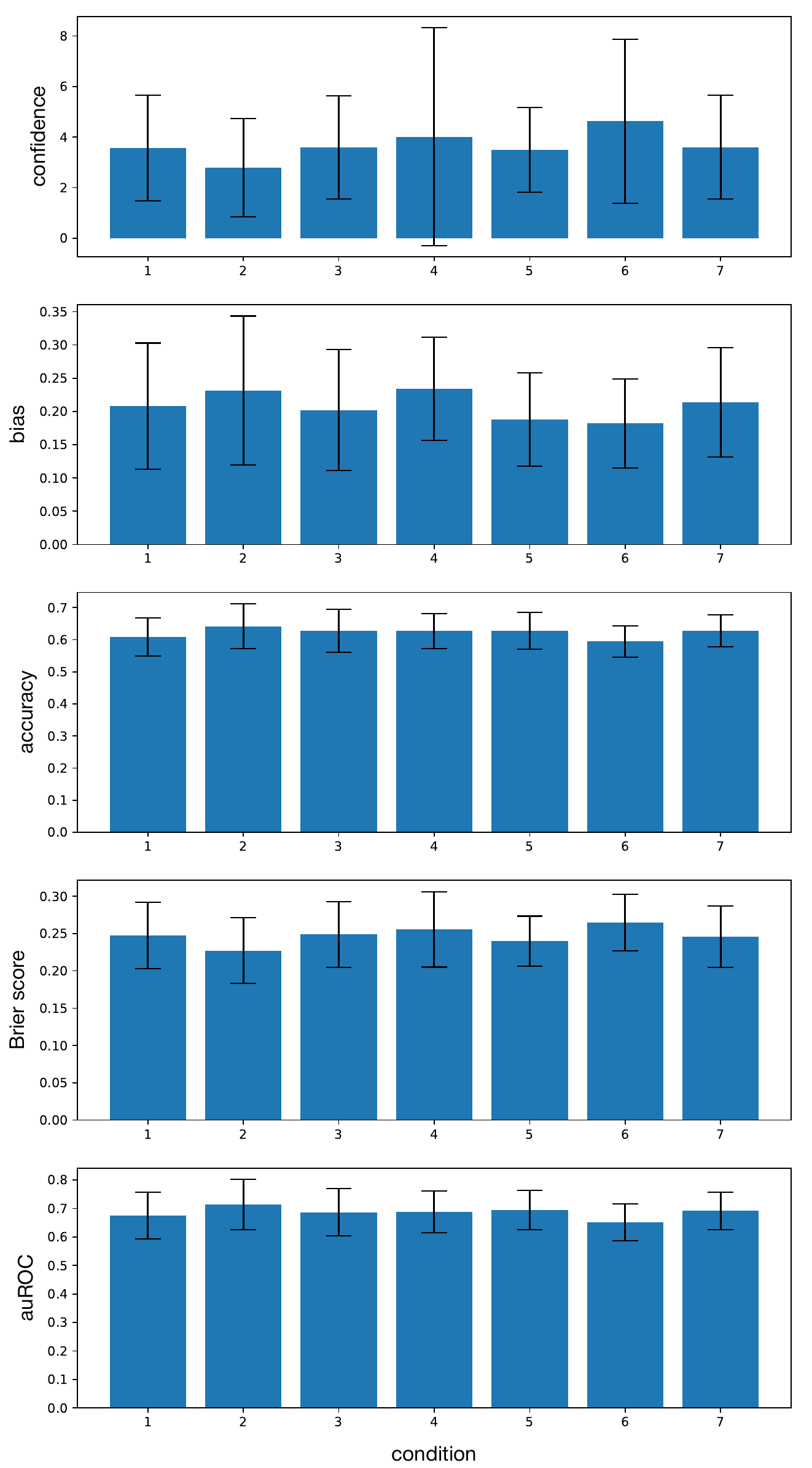}
    \caption{\footnotesize\textbf{Effect of feedback condition on individual prediction performance.} Participant confidence, bias, and probability-judgment performance metrics compared across different feedback conditions. Error bars represent standard deviation across subjects. The difference in confidence levels across all feedback conditions is not significant (permutation F-test, 10000 repetitions, $p>.05$). The only significant pairwise difference is between feedback conditions 2 and 6 (Table \ref{table:conditions}), $p<.005$ after Bonferroni correction for multiple pairwise tests. We performed a condition-label permutation test for each pair of conditions.}
     \label{fig:fig6}
\end{figure}

\begin{figure}
    \centering
    \includegraphics[width=\textwidth]{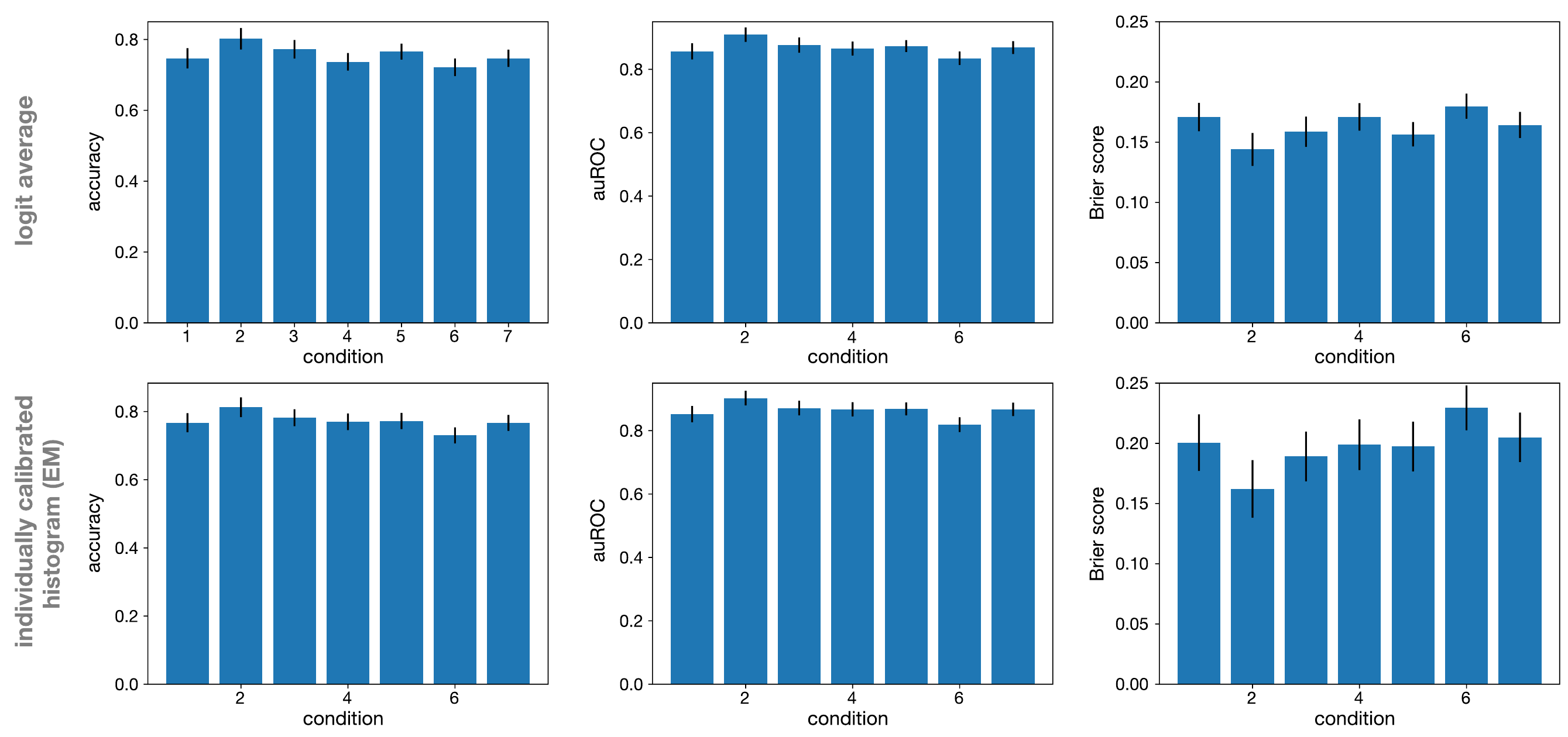}
    \caption{\footnotesize\textbf{Effect of feedback condition on collective inference performance.} Comparison of performance metrics across different feedback conditions for aggregate estimates. There was no significant effect of condition (7 conditions differing in feedback and reward) on any of the three performance metrics (accuracy, auROC, Brier score). Inference relied on a permutation F-test (10,000 repetitions, $p>.05$ for each metric). We also performed all possible comparisons for pairs of conditions and found no significant difference for any pair of feedback conditions (separate permutation test for each pair of conditions, $p>.05$ after Bonferroni correction for multiple tests across all pairs of the seven conditions).}
    \label{fig:fig7}
\end{figure}

\begin{figure}
    \centering
    \includegraphics[width=\textwidth]{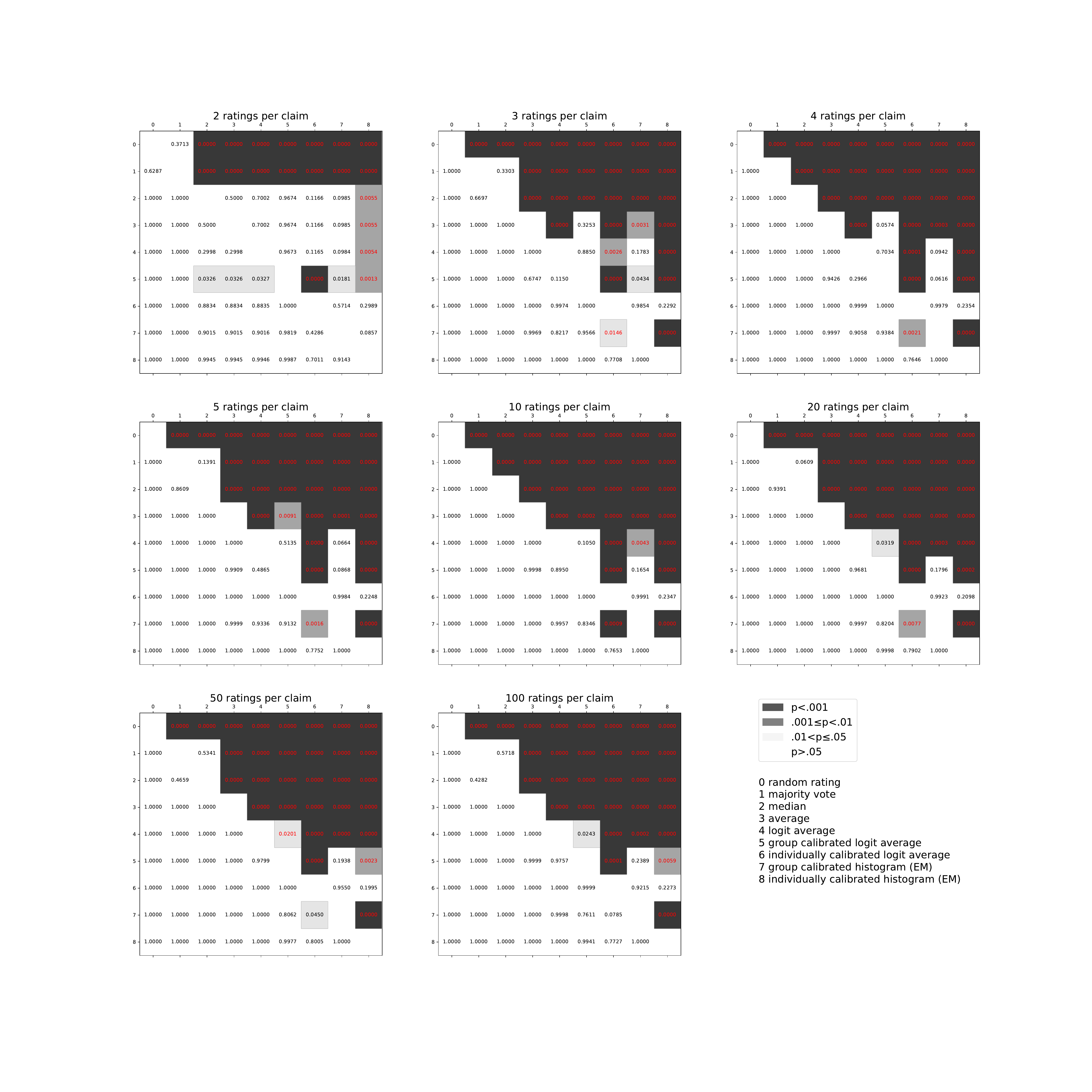}
    \caption{\footnotesize\textbf{Pairwise inferential comparisons among test accuracies of collective-inference algorithms presented in Fig. 2.} $p$-values for all pairwise comparisons in Fig. 2. $p$-values with text in red are statistically significant under the Benjamini-Hochberg procedure controlling the false discovery rate at level .05. The $p$-value reported in entry $(i,j)$ is a paired one-sided \textit{t}-test (374 degrees of freedom) testing if model $j$ is more accurate than model $i$.}
    \label{supp_fig:pairwise_comparison}
\end{figure}

We also analyzed how individual rating behavior depended on feedback and reward conditions. In particular, we evaluated how feedback and rewards affected confidence, bias, area under the receiver-operating characteristic, Brier score, and accuracy (Supplementary Figure \ref{fig:fig6}). We found no significant overall association between any of the five descriptors of rating behavior and the feedback and reward condition ($p>0.05$, permutation F-test comparing intra- and intergroup variance with 2-factor bootstrap, 10,000 draws). Trait- and ability-related interindividual differences in accuracy and calibration are more pronounced than differences caused by variation across our feedback and reward conditions. The lack of a significant effect of our experimental variation of feedback and rewards suggests that the instruction to judge probabilities invokes a cognitive process that is somewhat insensitive to the manipulations we implemented. Overall these results demonstrate that collective-inference algorithms can work robustly despite substantial variation across individuals in terms of accuracy and calibration as well as the subtler variation caused by context-dependent motivational factors.

\section*{Discussion}

Probability ratings enable people to share nuanced judgments of their degree of belief in propositions. They provide an attractive interface for crowd judgment systems because (1) they combine the efficiency of a single click on a scale and the nuance of a continuous rating, and (2) they are anchored in the normative framework of probabilistic inference, even if individual judgments require corrective calibration. Probability ratings will be useful wherever collections of claims are to be evaluated by groups of people.

We've shown here that (1) the wisdom of the crowd reflected in probability ratings can be leveraged even using naive aggregation methods like the majority vote, (2) continuous ratings support better collective inferences than binary responses, (3) more accurate collective inferences can be obtained by methods that combine the ratings according to the rules of probability, (4) further gains can be achieved if we account for individual accuracy and miscalibration, which does not require truth labels if we use a judgment-generative probabilistic model, and (5) supervised learning with truth labels can be used to ensure calibrated collective inferences and to enable collective inference when we have just one rating or a few ratings per person and therefore cannot model individual behavior.

\textbf{Modeling truth-conditional rating dependency.}
In addition to individual miscalibration, an ideal judgment-generative model should capture the truth-conditional dependency among the ratings. Such a model would account for the redundancy of the evidence reflected in the multiple ratings of a claim and could in principle achieve well-calibrated collective inferences even without truth labels. Methods for probability pooling that take dependency into account have been proposed \cite{trickBayesianClassifierFusion2022}. However, they require more data than will be available in many applications. Modeling rating dependency among peers is challenging in realistic scenarios with many peers and a sparse ratings matrix, where the number of claims rated by any two peers may be small or 0. In practice, calibrated collective inference can use the approach we take here of relying on a small number of truth-labeled ratings to correct miscalibration of the aggregate.

\textbf{Social media and the game of truth.}
In the context of social media, probability ratings could complement ``likes'', the dominant 1-click response modality. Whereas ``likes'' reflect emotional responses, probability ratings reflect knowledge and reasoning, promising a better basis for algorithmic amplification of messages. Liking and believing are independent emotional and cognitive dimensions of response, deserving separate elicitation. Consider, for example, a social media post that makes a depressing but true claim about an infectious disease. Judging the claim to have high probability (if we have reason to believe it) enables us to support its amplification, despite the fact that we do not like the content of the message. An intriguing question for future research is whether probability ratings can form the basis of a social game in which participants compete for accuracy and calibration rather than for popularity.

The social media context raises another issue for future work: What if there are bad actors in the community who submit false ratings with manipulative intent? Our judgment-generative model is able to capture a negative association of an individual's ratings with the truth and interpret the evidence accordingly, with ratings favoring a proposition from such an individual counting against it and ratings disfavouring a proposition counting in its favor. We therefore expect some robustness to a small portion of bad actors. However, sophisticated bad actors could contribute many reliable ratings to achieve good standing according to the model, putting them in a position to distort the evidence for particular claims they wish to target. Identifying such bad actors is a difficult problem that our methods here do not address. More generally, peers may be reliable on certain topics, but unreliable or untrustworthy on others. More sophisticated judgment-generative models could learn to characterize the space of claims and model the reliability of an individual as a function of the type of claim. Such methods would require a large amount of rating data in total and for each individual. The methods we describe are immediately useful for groups of peers acting in good faith.

\pagebreak

\section*{Methods}

\subsection*{Collective inference algorithms}
\textbf{Random rating.} A baseline estimate that avoids aggregating multiple probability ratings for a given claim is to simply pick a rating at random.

\textbf{Majority vote.}
Perhaps the simplest (and most common) method of judgment aggregation is majority vote. We count how many probability ratings are greater than 0.5 and how many are smaller than 0.5 (ignoring judgments that are exactly 0.5). The majority vote estimate is 1 if there are more ratings greater than 0.5 and 0 if there are more ratings smaller than 0.5. If the two counts match, a random tie break is performed. Peers' confidence is ignored in the majority vote, and all judgments have an equal impact towards the collective estimate. Since the collective inference is binary, it can be evaluated in terms of the accuracy, but not in terms of the auROC or the Brier score.


\textbf{Rating average.} The rating average is the arithmetic average of all ratings of a claim. The confidence of a peer's probability judgment, thus, has an impact on the collective inference, with more extreme ratings influencing the collective inference more.

\textbf{Median rating.} The median rating is the median of all ratings of a claim. When the majority vote is 1, the median rating will be above 0.5. However, see section \textit{Idiosyncrasies of random ratings, majority vote, and median rating} in the Supplementary information for some subtleties.

\textbf{Most accurate peer.} Given a training set, we can sort peers according to each peer's rating accuracy on the training set. The aggregate estimate for a claim in the test set is the rating from the peer who had the highest accuracy on the training set (with random tie-break in case more than one peer achieved the highest accuracy).

\textbf{Logit sum and average.} If we assume that all peers are independent and that the prior probability of a claim being true or false is equal, then the probability that claim $i$ is true ($t_i=1$, where $\textbf{t}$ is a vector of size $M$ of binary truth values) given a set of peer probability judgments $\mathbf{r}_i=[r_{i1},r_{i2},...,r_{iN}]$ is:
\begin{align}
    p(t_i=1|\mathbf{r}_i)&=\frac{\prod_{j=1}^Np(t_i=1|r_{ij})}{\prod_{j=1}^Np(t_i=0|r_{ij})+\prod_{j=1}^Np(t_i=1|r_{ij})}\\
    &=\sigma\left(\sum_{j=1}^N\log p(t_i=1|r_{ij})-\log p(t_i=0|r_{ij})\right)
\end{align}
where $\sigma$ is the logistic sigmoid (expit) function. This is also known as the independent opinion pool \cite{bergerStatisticalDecisionTheory1985}. If we further assume that peers' probability judgments are well-calibrated, we have
\begin{equation}
    p(t_i=1|\mathbf{r}_i)=\sigma\left(\sum_{j=1}^N\text{logit}(r_{ij})\right).
\end{equation}
In reality, ratings are not independent because peers draw from overlapping evidence pools, so this estimate is generally overconfident.



\textbf{Calibrated logit sum and average.}
It is known that human probability ratings are not well-calibrated \cite{mellersIdentifyingCultivatingSuperforecasters2015,albrechtSimilarityupdatingModelProbability2021,brennerOverconfidenceProbabilityFrequency1996,pickhardtStudyPerformanceSubjective1974}. Given a training set of truth-labeled claims and a peer $j$'s corresponding probability ratings, we can learn the peer's confidence $c_j$ and bias $b_j$ using logistic regression:
\begin{equation}
    p(t_i=1|r_{ij}) = \sigma(\logit(r_{ij})/c_j-b_j).
\end{equation}
As before, the logits are summed if we assume each peer is independent. We call this model the individually-calibrated logit sum (or average if we average the logits). In general, we find estimates for these models are more stable if we average the logits. In particular, truth inferences do not necessarily become more extreme in confidence as the number $N$ of ratings from different peers increases.

We may choose instead to learn a global set of logistic regression parameters $c$ and $b$ by combining all peers' ratings and fitting a logistic regression model. We call these models group-calibrated.

\textbf{Judgement-generative model learned with expectation maximization.}
An alternative to variants of the independent opinion pool is to model the truth-conditional rating behavior of peers: $p(r_{ij}|t_i)$. On this basis, we can model the joint density over all peers' ratings of claim $i$: $p(\mathbf{r}_i|t_i)$. We assume peers are independent, so the joint density is the product of the individual peer rating densities. We can then use Bayes' rule to infer the probability of the claim:
\begin{equation}
    p(t_i=1|\mathbf{r}_i)=\frac{p(\mathbf{r}_i|t_i=1)\cdot p(t_i=1)}{p(\mathbf{r}_i|t_i=1)\cdot p(t_i=1)+p(\mathbf{r}_i|t_i=0)\cdot p(t_i=0)}.
\end{equation}

In order to learn the peer behavior models $p(r_{ij}|t_i),$ we can either use a training set with truth labels or infer them by using the Expectation-Maximization (EM) algorithm \cite{dempsterMaximumLikelihoodIncomplete1977} to fit peer parameters while estimating the probability of each claim, as is the general strategy in \cite{dawidMaximumLikelihoodEstimation1979}. Note that the EM algorithm can be also used as a form of semi-supervised learning by replacing its inferences for labeled data points with their corresponding truth labels.
\\
\\
The EM algorithm first calculates the posterior claim probabilities of each claim given the current estimate of user/claim traits (Expectation step) and then maximizes the expected value of the joint log-likelihood of the complete data (ratings and truth values) under the previously calculated posterior claim probabilities (Maximization step). Under our model, the EM objective function takes the form:
\begin{equation}
    Q(\theta|\theta^{\text{old}})=\mathbb{E}_{p(\mathbf{t}|\mathbf{R};\theta^{(\text{old})})}[\log p(\mathbf{R},\mathbf{t}|\theta)],
\end{equation}
where $\mathbf{t}$ is the vector of truth values and $\mathbf{R}=[\mathbf{r}_1,\mathbf{r}_2,...,\mathbf{r}_M]$ is the ratings matrix. If we assume the peers are independent and equal probability for true and false claims, we have
\begin{align}
    \argmax_\theta Q(\theta|\theta^{\text{old}})&=\argmax_\theta\mathbb{E}_{p(\mathbf{t}|\mathbf{R};\theta^{\text{old}})}\left[\sum_{i=1}^M\sum_{j\in J_i}\log p(r_{ij}|t_i;\theta_j)\right]\\
    &=\sum_{i=1}^M\sum_{j\in J_i}\pi_i\log p(r_{ij}|t_i=1;\theta_j)+(1-\pi_i)\log p(r_{ij}|t_i=0;\theta_j),
\end{align}
where $J_i$ is the set of indices of peers who rated claim $i$ and $\pi_i$ is defined as $p(t_i=1|r_{ij},\theta^{(\text{old})}).$
\\\\
Since each peer has her own set of parameters $\theta_j$, maximizing $Q(\theta|\theta^\text{old})$ splits into a set of subproblems, one for each user. We choose for our generative model $p(r_{ij}|t_i;\theta_j)$ a histogram of 5 evenly spaced bins over the unit interval (with similar results with different numbers of bins and uneven bin sizes). Under this model, the M-step has a closed form solution for each peer. Given a partitioning of the unit interval $0=g_1<g_2<\cdots<g_{B+1}=1$, where $B$ is the number of bins in our histogram, we have
\begin{equation}
    p(g_l\leq r_{ij}<g_{l+1}|t_i=1)\propto\sum_{i'\in I_j}^M \pi_{i'}\mathbb{I}(g_l\leq r_{i'j}<g_{l+1}),\; l=1,...,B.
\end{equation}
If the bin spacing is uniform, the normalizing constant is $B\sum_{i\in I_j}\pi_i$. Replacing $\pi_{i'}$ with $1-\pi_{i'}$ in the above gives the result for the False-conditional histogram generative model.
\\
\\
Given the peer parameters and a group of ratings of a particular claim, we can infer the probability that the claim is true as follows:
\begin{align}
    p(t_i=1|r_{i,J_i};\theta_{J_i}) &= \frac{p(r_{i,J_i}|t_i=1;\theta_{J_i})}
    {p(r_{i,J_i}|t_i=1;\theta_{J_i})+p(r_{i,J_i}|t_i=0;\theta_{J_i})}\\
    &=\sigma\left(\log\frac{p(r_{i,J_i}|t_i=1;\theta_{J_i})}{p(r_{i,J_i}|t_i=0;\theta_{J_i})}\right).
\end{align}
If the users' ratings are independent, we have:
\begin{equation}
    p(t_i=1|r_{i,J_i};\theta_{J_i})=\sigma\left(\sum_{j\in J_i}\log\frac{p(r_{i,j}|t_i=1;\theta_j)}{p(r_{i,j}|t_i=0;\theta_j)}\right).
\end{equation}

The inference, thus, involves summing estimates of $\log [p(r_{ij}|t_i=1;\theta_j)/p(r_{ij}|t_i=0;\theta_j)]$ across peers, where $p(r_{ij}|t_i=1;\theta_j)$ and $p(r_{ij}|t_i=0;\theta_j)$ are provided by our generative model with peer-specific parameters $\theta_j$. Because of the assumption of truth-conditionally independent ratings, the collective inferences are expected to be overconfident. To avoid modeling the truth-conditional rating dependencies, we propose to take a supervised recalibration approach, which requires a small number of truth-labeled claims (e.g. 100). Using the truth labels, we can recalibrate our collective-inference logit estimates using the same approach as on the level of individual peer ratings: by fitting a bias and confidence parameter, which amounts to training a logistic regression model. 

\textbf{Dawid-Skene.}
The rating data is binarized by applying a Heaviside function $H(\cdot)$ after subtracting 0.5 from each rating. This maps ratings larger than 0.5 to 1 and ratings smaller than 0.5 to 0. We then fit the two-coin Dawid Skene model \cite{dawidMaximumLikelihoodEstimation1979}. The Dawid-Skene model defines the probability that each user gives the correct (binarized) rating to a claim: $p(H(r_{ij}-.5)=1|t=1)=\theta^{(1)}_j$ and $p(H(r_{ij}-.5)=0|t=0)=\theta^{(0)}_j$. As with the histogram generative model, we set the prior on $t$ to be flat and we use EM to estimate $\theta_j^{(1)}$ and $\theta_j^{(0)}.$ Again, the M-step has a closed form solution:
\begin{gather}
    \theta_j^{(1)}=\frac{\sum_{i\in I_j}H(r_{ij}-.5)\pi_i}{\sum_{i\in I_j}\pi_i}\\
    \theta_j^{(0)}=\frac{\sum_{i\in I_j}(1-H(r_{ij}-.5))(1-\pi_i)}{\sum_{i\in I_j}1-\pi_i},
\end{gather}
where $I_j$ is the indices of the claims that peer $j$ has rated.

\textbf{Supervised centered-moment logistic regression.}
Ideally, we would be able to learn a function that maps from the distribution of peer ratings of a particular claim to an estimate of the probability of the claim. The distribution of ratings a claim has received can be characterized by its centered moments. This approach does not require a large number of ratings. The $m$-th centered empirical moment of the set of ratings for claim $i$ is:
\begin{equation}
    \frac{1}{|J_i|}\sum_{j\in J_i}(r_{i,j}-\mu_{i,J_i})^m,
\end{equation}
where $|\cdot|$ is the cardinality operator and $\mu_{i,J_i}:=\frac{1}{|J_i|}\sum_{j\in J_i}r_{i,j}.$

We characterize the distribution of ratings by 5 real numbers: The first raw moment (mean), the second centered moment (variance), the third centered moment (skew), the fourth centered moment (kurtosis), and the fifth centered moment. The centered-moment logistic regression model fits a weight to each of the five moments of the empirical distribution of ratings in a labeled training set. The weighted combination is passed through the standard logistic  function to provide the probability estimate $p(t_i=1|r_{i,J_i}).$ This model can also be trained on binarized ratings.


\subsection*{Behavioral experiment}
We created a set of 1,200 general knowledge claims equally partitioned into six categories, giving 200 claims per category: history, geography, science, social sciences and politics, sports and leisure, and arts and entertainment. Each category had an equal number of true and false claims. Our full list of claims and truth values is provided in the data repository as detailed in the Data availability statement. 

Recognizing that the baseline knowledge assumed in this study centered predominantly on U.S. contexts, we sought participants through www.prolific.org who self-identified as U.S. citizens. We further restricted the participant pool to those who had no rejections from previous studies and at least 20 studies completed. Prolific users who met these criteria could begin participation in the study independently and were included on a first-come, first-served basis. The 1,200 claims were augmented by 10 trivially easy claims (e.g. ``All fish can fly") to gauge user engagement. We excluded participants who failed on these claims. Each participant was asked to judge all 1,200 claims across six sessions on different days. Of 504 paid participants, 376 completed all six sessions and only these were retained in the data set. Given the selection process, our sample is thus not representative of either the human population or the U.S. population. However, the remaining group of 376 participants was diverse in age (18 to 65 years, median: 25 years) and gender (61\% female, 5\% nonbinary, 34\% male), and to a lesser extent in race (7\% Asian, 6\% Black, 1\% Native American, 1\% Pacific Islander, 82\% White; 3\% Prefer not to say).

We used www.meadows-research.com as the platform to host our studies. Participants were instructed on how to rate claims with a few trial examples provided before the study began. Each user completed 6 studies, each consisting of a random portion of 200 of the 1,200 claims. The 200 claims were further split into 4 blocks of 50 claims. Users were allowed to begin each study at their own pace. Once a study began, each claim had to be completed within 20 seconds. Each of the included 376 participants judged each of the 1,200 claims.

We randomly assigned each user one of seven different feedback conditions, shown in Table \ref{table:conditions}.
\begin{table}[h!]
\begin{tabular}{ ll } 
Condition & Description\\
 \hline
 1 & No feedback\\
 \hline
 2 & True/False feedback for each claim presented as a list after a block of trials\\
 \hline
 3 & True/False feedback after each trial\\ 
 \hline
 4 & True/False feedback after each trial and \% of users whose rating is correct\\ 
 \hline
 5 & True/False feedback after each trial and reward $\propto|r-t|$\\ 
 \hline
 6 & True/False feedback after each trial and reward $\propto|r-t|^2$\\ 
 \hline
 7 & True/False feedback after each trial and reward $\propto|r-t|^3$\\ 
 \hline
\end{tabular}
\caption{Feedback conditions.}
\label{table:conditions}
\end{table}
In condition 1, participants received no feedback or reward. In all other conditions, participants received True/False feedback. In conditions 5, 6, and 7, participants additionally received imaginary coin rewards for accurate judgments. The reward fell off with the discrepancy $|r-t|$ between the rating $r\in [0,1]$ and the truth $t\in\{0,1\}$. Each of conditions 5, 6, and 7 used a different reward function, encouraging overconfident rating (reward $\propto |r-t|$), well-calibrated rating (reward $\propto |r-t|^2$, proper scoring rule), and underconfident rating (reward $\propto |r-t|^3$), respectively. These conditions were included to enable us to gauge the robustness of the collective-inference algorithms to varying incentives that might distort the human judgments.

\subsection*{Data partitioning for training and testing}
The performance measures we report are from a held out test set with the trainable models trained on a disjoint training set. For each bootstrap resample of the data, we randomly partition the data into $K$ equal bins and leave each bin out as a test set and train on the other $K-1$ bins, giving $K$ performance metrics, which we average together. 

\subsection*{Data-based simulation of sparse rating matrices}
In real-world applications, we will not have a dense matrix of probability ratings as acquired in our online behavioral experiment.  For peers in a social network rating claims they encounter, for example, we expect that every claim is rated by a small subset of the peers, and that every peer rates a small subset of the claims. We therefore create sparse rating matrices by resampling, so as to compare the performance of different collective-inference algorithms as a function of the number of ratings per claim and the number of ratings per peer.

We use a resampling method that we call ``thinning" to obtain sparse ratings matrices as may be encountered in practice. Thinning the matrix by factor $k$ along the peers dimension involves replacing each peer's column by $k$ copies of that column. For each peer and row, we then retain only one of the $k$ identical ratings, choosing the one to retain at random. The other copies are set to ``missing". The new matrix contains the same ratings as the original matrix, but the ratings appear to originate from $k$ times as many peers and only a fraction of $1/k$ of all possible ratings is present. Although the sparsified matrix contains the same set of ratings, it provides less information to a collective-inference algorithm because it does not specify which of the ratings in different columns actually came from the same peer. An algorithm like histogram-EM must try to infer more peer models given fewer ratings per peer. 

The same thinning technique is applied to the claims, replacing each claim's row with $l$ copies of that row. For each present original rating, we retain only one of the $l$ identical copies (choosing which to retain at random). 
Expanding a dense $M$ by $N$ matrix into a sparse $M \cdot l$ by $N \cdot k$ matrix preserves all ratings, while simulating a larger number of peers ($N \cdot k$) and claims ($M \cdot l$). Subsampling further enables us to control the number of ratings per claim and peer. 

\subsection*{Statistical inference for comparing collective-inference methods}
Estimates of the accuracy, area under the receiver operating characteristic (auROC) and Brier score of different collective-inference algorithms are affected by measurement noise as well as variation due to the sample of participants and claims. We are interested in statistical inferences that hold, not just for our sample of participants and our 1,200 claims, but for the underlying populations that our participants and claims can be considered random samples from (e.g. U.S. citizen prolific users; see section \textit{Behavioral experiment}, above, for limitations of our samples of participants and claims). This motivates a conservative approach to frequentist statistical inference in which we treat both participants and claims as random effects. For statistical comparisons among inference methods, we therefore simulate the variation due to sampling of claims and participants by resampling both participants and claims with replacement (two-factor bootstrap) \cite{schuttStatisticalInferenceRepresentational2023}.


The two-factor bootstrap provides variance estimates for each performance measure (accuracy, auROC, Brier score) for each collective-inference algorithm as well as variance estimates for the difference in these performance measures for each pair of collective-inference algorithms. For each performance measure and pair of collective-inference algorithms, we compute a performance difference for each two-factor bootstrap sample. The variance of these differences forms the basis for inference using paired $t$-tests. The number of degrees of freedom is set to the smaller of the two numbers of samples (number of participants or number of claims) minus 1. 

In Fig. \ref{fig:fig2}, we are interested in the degree to which each step in a progression from naive to sophisticated collective-inference algorithms improves performance as a function of the number of ratings per claim. We therefore perform single-tailed $t$-tests, testing for a performance improvement for each step. We use a Bonferroni correction to account for $9$ comparisons (for different numbers of ratings per claim). In Fig. \ref{fig:fig4}, we use a paired \textit{t}-test on the performance difference between two models using a two-factor bootstrap to estimate variances. We again use a Bonferroni correction to account for the $4$ statistical tests done for each algorithm comparison across the four panels, corresponding to different numbers of truth labels.

\subsection*{Data availability}

The data set of 451,200 probability ratings (of 1,200 claims by each of the 376 peers), the 1,200 claims (falling in 6 topic categories) and their truth labels will be shared with the community in an open-science repository upon journal publication of the paper. 

\subsection*{Code availability}
Our code repository is available to reviewers now and will be shared on GitHub upon journal publication.

\pagebreak
\printbibliography

@book{aggarwalRecommenderSystemsTextbook2016,
  title = {Recommender {{Systems}}: {{The Textbook}}},
  author = {Aggarwal, Charu C.},
  year = {2016},
  edition = {1st ed.},
  publisher = {Springer},
  address = {Cham Heidelberg New York Dordrecht London}
}

@article{albrechtSimilarityupdatingModelProbability2021,
  title = {The Similarity-Updating Model of Probability Judgment and Belief Revision},
  author = {Albrecht, Rebecca and Jenny, Mirjam A. and Nilsson, H{\aa}kan and Rieskamp, J{\"o}rg},
  year = {2021},
  journal = {Psychological Review},
  volume = {128},
  number = {6},
  pages = {1088--1111},
  publisher = {US: American Psychological Association}
}

@article{almaatouqAdaptiveSocialNetworks2020,
  title = {Adaptive Social Networks Promote the Wisdom of Crowds},
  author = {Almaatouq, Abdullah and {Noriega-Campero}, Alejandro and Alotaibi, Abdulrahman and Krafft, P. M. and Moussaid, Mehdi and Pentland, Alex},
  year = {2020},
  journal = {Proceedings of the National Academy of Sciences},
  volume = {117},
  number = {21},
  pages = {11379--11386},
  doi = {10.1073/pnas.1917687117}
}

@book{bailBreakingSocialMedia2021,
  title = {Breaking the Social Media Prism: How to Make Our Platforms Less Polarizing},
  shorttitle = {Breaking the Social Media Prism},
  author = {Bail, Christopher},
  year = {2021},
  publisher = {Princeton University Press},
  address = {Princeton},
  isbn = {978-0-691-20342-3},
  langid = {english},
  lccn = {HM742 .B37 2021}
}

@article{bailExposureOpposingViews2018,
  title = {Exposure to Opposing Views on Social Media Can Increase Political Polarization},
  author = {Bail, Christopher A. and Argyle, Lisa P. and Brown, Taylor W. and Bumpus, John P. and Chen, Haohan and Hunzaker, M. B. Fallin and Lee, Jaemin and Mann, Marcus and Merhout, Friedolin and Volfovsky, Alexander},
  year = {2018},
  journal = {Proceedings of the National Academy of Sciences},
  volume = {115},
  number = {37},
  pages = {9216--9221}
}

@book{bergerStatisticalDecisionTheory1985,
  title = {Statistical {{Decision Theory}} and {{Bayesian Analysis}}},
  author = {Berger, James O.},
  year = {1985},
  series = {Springer {{Series}} in {{Statistics}}},
  publisher = {Springer},
  address = {New York, NY},
  doi = {10.1007/978-1-4757-4286-2}
}

@article{brennerOverconfidenceProbabilityFrequency1996,
  title = {Overconfidence in Probability and Frequency Judgments: {{A}} Critical Examination},
  shorttitle = {Overconfidence in {{Probability}} and {{Frequency Judgments}}},
  author = {Brenner, Lyle A. and Koehler, Derek J. and Liberman, Varda and Tversky, Amos},
  year = {1996},
  journal = {Organizational Behavior and Human Decision Processes},
  volume = {65},
  number = {3},
  pages = {212--219},
  doi = {10.1006/obhd.1996.0021}
}

@article{chaseVisionsRationality1998,
  title = {Visions of Rationality},
  author = {Chase, Valerie M and Hertwig, Ralph and Gigerenzer, Gerd},
  year = {1998},
  journal = {Trends in Cognitive Sciences},
  volume = {2},
  number = {6},
  pages = {206--214},
  doi = {10.1016/S1364-6613(98)01179-6}
}

@article{clemenCombiningForecastsReview1989,
  title = {Combining Forecasts: {{A}} Review and Annotated Bibliography},
  shorttitle = {Combining Forecasts},
  author = {Clemen, Robert T.},
  year = {1989},
  journal = {International Journal of Forecasting},
  volume = {5},
  number = {4},
  pages = {559--583},
  copyright = {https://www.elsevier.com/tdm/userlicense/1.0/}
}

@article{dawidMaximumLikelihoodEstimation1979,
  title = {Maximum Likelihood Estimation of Observer Error-Rates Using the {{EM}} Algorithm},
  author = {Dawid, A. P. and Skene, A. M.},
  year = {1979},
  journal = {Journal of the Royal Statistical Society. Series C (Applied Statistics)},
  volume = {28},
  number = {1},
  pages = {20--28},
  doi = {10.2307/2346806}
}

@article{degrootReachingConsensus1974,
  title = {Reaching a Consensus},
  author = {Degroot, Morris H.},
  year = {1974},
  journal = {Journal of the American Statistical Association},
  volume = {69},
  number = {345},
  pages = {118--121},
  doi = {10.1080/01621459.1974.10480137}
}

@article{delvicarioSpreadingMisinformationOnline2016,
  title = {The Spreading of Misinformation Online},
  author = {Del Vicario, Michela and Bessi, Alessandro and Zollo, Fabiana and Petroni, Fabio and Scala, Antonio and Caldarelli, Guido and Stanley, H. Eugene and Quattrociocchi, Walter},
  year = {2016},
  journal = {Proceedings of the National Academy of Sciences},
  volume = {113},
  number = {3},
  pages = {554--559},
  doi = {10.1073/pnas.1517441113}
}

@article{dempsterMaximumLikelihoodIncomplete1977,
  title = {Maximum Likelihood from Incomplete Data via the {{{\emph{EM}}}} Algorithm},
  author = {Dempster, A. P. and Laird, N. M. and Rubin, D. B.},
  year = {1977},
  journal = {Journal of the Royal Statistical Society: Series B (Methodological)},
  volume = {39},
  number = {1},
  pages = {1--22},
  doi = {10.1111/j.2517-6161.1977.tb01600.x}
}

@article{dietrichProbabilisticOpinionPooling2017,
  title = {Probabilistic Opinion Pooling Generalized. {{Part}} One: {{General}} Agendas},
  shorttitle = {Probabilistic Opinion Pooling Generalized. {{Part}} One},
  author = {Dietrich, Franz and List, Christian},
  year = {2017},
  journal = {Social Choice and Welfare},
  volume = {48},
  number = {4},
  pages = {747--786},
  doi = {10.1007/s00355-017-1034-z}
}

@book{embretsonItemResponseTheory2013,
  title = {Item {{Response Theory}}},
  author = {Embretson, Susan E. and Reise, Steven P.},
  year = {2013},
  publisher = {Psychology Press},
  doi = {10.4324/9781410605269}
}

@article{galtonVoxPopuli1907,
  title = {Vox {{Populi}}},
  author = {Galton, Francis},
  year = {1907},
  journal = {Nature},
  volume = {75},
  number = {1949},
  pages = {450--451},
  doi = {10.1038/075450a0}
}

@article{geersOnlineMisinformationEngagement2024,
  title = {The {{Online Misinformation Engagement Framework}}},
  author = {Geers, Michael and {Swire-Thompson}, Briony and {Lorenz-Spreen}, Philipp and Herzog, Stefan M. and Kozyreva, Anastasia and Hertwig, Ralph},
  year = {2024},
  journal = {Current Opinion in Psychology},
  volume = {55},
  pages = {101739}
}

@article{genestCombiningProbabilityDistributions1986,
  title = {Combining Probability Distributions: {{A}} Critique and an Annotated Bibliography},
  shorttitle = {Combining {{Probability Distributions}}},
  author = {Genest, Christian and Zidek, James V.},
  year = {1986},
  journal = {Statistical Science},
  volume = {1},
  number = {1},
  pages = {114--135}
}

@article{gershmanComputationalRationalityConverging2015,
  title = {Computational Rationality: {{A}} Converging Paradigm for Intelligence in Brains, Minds, and Machines},
  shorttitle = {Computational Rationality},
  author = {Gershman, S. J. and Horvitz, E. J. and Tenenbaum, J. B.},
  year = {2015},
  journal = {Science},
  volume = {349},
  number = {6245},
  pages = {273--278},
  doi = {10.1126/science.aac6076}
}

@article{griffithsOptimalPredictionsEveryday2006,
  title = {Optimal {{Predictions}} in {{Everyday Cognition}}},
  author = {Griffiths, Thomas L. and Tenenbaum, Joshua B.},
  year = {2006},
  journal = {Psychological Science},
  volume = {17},
  number = {9},
  pages = {767--773},
  doi = {10.1111/j.1467-9280.2006.01780.x}
}

@article{griffithsProbabilisticModelsCognition2010,
  title = {Probabilistic Models of Cognition: {{Exploring}} Representations and Inductive Biases},
  shorttitle = {Probabilistic Models of Cognition},
  author = {Griffiths, Thomas L. and Chater, Nick and Kemp, Charles and Perfors, Amy and Tenenbaum, Joshua B.},
  year = {2010},
  journal = {Trends in Cognitive Sciences},
  volume = {14},
  number = {8},
  pages = {357--364},
  doi = {10.1016/j.tics.2010.05.004}
}

@book{hambletonItemResponseTheory1985,
  title = {Item {{Response Theory}}},
  author = {Hambleton, Ronald K. and Swaminathan, Hariharan},
  year = {1985},
  publisher = {Springer Netherlands},
  address = {Dordrecht},
  doi = {10.1007/978-94-017-1988-9}
}

@book{haneaExpertJudgementRisk2021,
  title = {Expert {{Judgement}} in {{Risk}} and {{Decision Analysis}}},
  author = {Hanea, Anca M. and Nane, Gabriela F. and Bedford, Tim and French, Simon},
  year = {2021},
  publisher = {Springer Nature}
}

@article{harveyConfidenceJudgment1997,
  title = {Confidence in Judgment},
  author = {Harvey, Nigel},
  year = {1997},
  journal = {Trends in Cognitive Sciences},
  volume = {1},
  number = {2},
  pages = {78--82},
  copyright = {https://www.elsevier.com/tdm/userlicense/1.0/}
}

@article{hertwigDecisionsExperienceEffect2004,
  title = {Decisions {{From Experience}} and the {{Effect}} of {{Rare Events}} in {{Risky Choice}}},
  author = {Hertwig, Ralph and Barron, Greg and Weber, Elke U and Erev, Ido},
  year = {2004},
  journal = {Psychological Science},
  volume = {15},
  number = {8},
  pages = {534--539}
}

@article{hertwigFastFrugalHeuristics2009,
  title = {Fast and Frugal Heuristics: {{Tools}} of Social Rationality},
  shorttitle = {Fast and {{Frugal Heuristics}}},
  author = {Hertwig, Ralph and Herzog, Stefan M.},
  year = {2009},
  journal = {Social Cognition},
  volume = {27},
  number = {5},
  pages = {661--698},
  doi = {10.1521/soco.2009.27.5.661}
}

@book{keeneyDecisionsMultipleObjectives1976,
  title = {Decisions with {{Multiple Objectives}}},
  author = {Keeney, Ralph and Raiffa, Howard},
  year = {1976},
  publisher = {New York: Wiley}
}

@article{kerrGroupPerformanceDecision2004,
  title = {Group Performance and Decision Making},
  author = {Kerr, Norbert L. and Tindale, R. Scott},
  year = {2004},
  journal = {Annual Review of Psychology},
  volume = {55},
  number = {1},
  pages = {623--655},
  doi = {10.1146/annurev.psych.55.090902.142009}
}

@article{klaymanOverconfidenceItDepends1999,
  title = {Overconfidence: {{It}} Depends on How, What, and Whom You Ask},
  author = {Klayman, Joshua and Soll, Jack B. and {Gonz{\'a}lez-Vallejo}, Claudia and Barlas, Sema},
  year = {1999},
  journal = {Organizational Behavior and Human Decision Processes},
  volume = {79},
  number = {3},
  pages = {216--247},
  doi = {10.1006/obhd.1999.2847}
}

@article{kozyrevaCitizensInternetConfronting2020,
  title = {Citizens versus the Internet: {{Confronting}} Digital Challenges with Cognitive Tools},
  shorttitle = {Citizens {{Versus}} the {{Internet}}},
  author = {Kozyreva, Anastasia and Lewandowsky, Stephan and Hertwig, Ralph},
  year = {2020},
  journal = {Psychological Science in the Public Interest},
  volume = {21},
  number = {3},
  pages = {103--156},
  doi = {10.1177/1529100620946707}
}

@article{lazerScienceFakeNews2018,
  title = {The Science of Fake News},
  author = {Lazer, David M. J. and Baum, Matthew A. and Benkler, Yochai and Berinsky, Adam J. and Greenhill, Kelly M. and Menczer, Filippo and Metzger, Miriam J. and Nyhan, Brendan and Pennycook, Gordon and Rothschild, David and Schudson, Michael and Sloman, Steven A. and Sunstein, Cass R. and Thorson, Emily A. and Watts, Duncan J. and Zittrain, Jonathan L.},
  year = {2018},
  journal = {Science},
  volume = {359},
  number = {6380},
  pages = {1094--1096},
  doi = {10.1126/science.aao2998}
}

@article{leeInferringExpertiseKnowledge2012,
  title = {Inferring Expertise in Knowledge and Prediction Ranking Tasks},
  author = {Lee, Michael D. and Steyvers, Mark and {de Young}, Mindy and Miller, Brent},
  year = {2012},
  journal = {Topics in Cognitive Science},
  volume = {4},
  number = {1},
  pages = {151--163},
  doi = {10.1111/j.1756-8765.2011.01175.x}
}

@article{lorenz-spreenHowBehaviouralSciences2020,
  title = {How Behavioural Sciences Can Promote Truth, Autonomy and Democratic Discourse Online},
  author = {{Lorenz-Spreen}, Philipp and Lewandowsky, Stephan and Sunstein, Cass R. and Hertwig, Ralph},
  year = {2020},
  journal = {Nature Human Behaviour},
  volume = {4},
  number = {11},
  pages = {1102--1109},
  doi = {10.1038/s41562-020-0889-7}
}

@article{maBayesianDecisionModels2019,
  title = {Bayesian Decision Models: {{A}} Primer},
  shorttitle = {Bayesian Decision Models},
  author = {Ma, Wei Ji},
  year = {2019},
  journal = {Neuron},
  volume = {104},
  number = {1},
  pages = {164--175},
  publisher = {Elsevier}
}

@article{mellersIdentifyingCultivatingSuperforecasters2015,
  title = {Identifying and {{Cultivating Superforecasters}} as a {{Method}} of {{Improving Probabilistic Predictions}}},
  author = {Mellers, Barbara and Stone, Eric and Murray, Terry and Minster, Angela and Rohrbaugh, Nick and Bishop, Michael and Chen, Eva and Baker, Joshua and Hou, Yuan and Horowitz, Michael and Ungar, Lyle and Tetlock, Philip},
  year = {2015},
  journal = {Perspectives on Psychological Science},
  volume = {10},
  number = {3},
  pages = {267--281}
}

@article{muchnikSocialInfluenceBias2013,
  title = {Social Influence Bias: {{A}} Randomized Experiment},
  shorttitle = {Social {{Influence Bias}}},
  author = {Muchnik, Lev and Aral, Sinan and Taylor, Sean J.},
  year = {2013},
  journal = {Science},
  volume = {341},
  number = {6146},
  pages = {647--651},
  doi = {10.1126/science.1240466}
}

@article{ohaganExpertKnowledgeElicitation2019,
  title = {Expert Knowledge Elicitation: {{Subjective}} but Scientific},
  shorttitle = {Expert {{Knowledge Elicitation}}},
  author = {O'Hagan, Anthony},
  year = {2019},
  journal = {The American Statistician},
  volume = {73},
  number = {sup1},
  pages = {69--81},
  doi = {10.1080/00031305.2018.1518265}
}

@article{pennycookFightingMisinformationSocial2019,
  title = {Fighting Misinformation on Social Media Using Crowdsourced Judgments of News Source Quality},
  author = {Pennycook, Gordon and Rand, David G.},
  year = {2019},
  journal = {Proceedings of the National Academy of Sciences},
  volume = {116},
  number = {7},
  pages = {2521--2526},
  doi = {10.1073/pnas.1806781116}
}

@article{pennycookPsychologyFakeNews2021,
  title = {The {{Psychology}} of {{Fake News}}},
  author = {Pennycook, Gordon and Rand, David G.},
  year = {2021},
  journal = {Trends in Cognitive Sciences},
  volume = {25},
  number = {5},
  pages = {388--402}
}

@article{pickhardtStudyPerformanceSubjective1974,
  title = {A Study of the Performance of Subjective Probability Assessors},
  author = {Pickhardt, Robert C. and Wallace, John B.},
  year = {1974},
  journal = {Decision Sciences},
  volume = {5},
  number = {3},
  pages = {347--363},
  doi = {10.1111/j.1540-5915.1974.tb00621.x}
}

@article{raykarLearningCrowds2010,
  title = {Learning {{From Crowds}}},
  author = {Raykar, Vikas C. and Yu, Shipeng and Zhao, Linda H. and Valadez, Gerardo Hermosillo and Florin, Charles and Bogoni, Luca and Moy, Linda},
  year = {2010},
  journal = {Journal of Machine Learning Research},
  volume = {11},
  number = {43},
  pages = {1297--1322}
}

@inproceedings{resnickGroupLensOpenArchitecture1994,
  title = {{{GroupLens}}: An Open Architecture for Collaborative Filtering of Netnews},
  shorttitle = {{{GroupLens}}},
  booktitle = {Proceedings of the 1994 {{ACM}} Conference on {{Computer}} Supported Cooperative Work},
  author = {Resnick, Paul and Iacovou, Neophytos and Suchak, Mitesh and Bergstrom, Peter and Riedl, John},
  year = {1994},
  series = {{{CSCW}} '94},
  pages = {175--186},
  publisher = {Association for Computing Machinery},
  address = {New York, NY, USA}
}

@inproceedings{sarwarItembasedCollaborativeFiltering2001,
  title = {Item-Based Collaborative Filtering Recommendation Algorithms},
  booktitle = {Proceedings of the 10th International Conference on {{World Wide Web}}},
  author = {Sarwar, Badrul and Karypis, George and Konstan, Joseph and Riedl, John},
  year = {2001},
  series = {{{WWW}} '01},
  pages = {285--295},
  publisher = {Association for Computing Machinery},
  address = {New York, NY, USA}
}

@article{schmidtAnatomyNewsConsumption2017,
  title = {Anatomy of News Consumption on {{Facebook}}},
  author = {Schmidt, Ana Luc{\'i}a and Zollo, Fabiana and Del Vicario, Michela and Bessi, Alessandro and Scala, Antonio and Caldarelli, Guido and Stanley, H. Eugene and Quattrociocchi, Walter},
  year = {2017},
  journal = {Proceedings of the National Academy of Sciences},
  volume = {114},
  number = {12},
  pages = {3035},
  doi = {10.1073/pnas.1617052114}
}

@article{schuttStatisticalInferenceRepresentational2023,
  title = {Statistical Inference on Representational Geometries},
  author = {Sch{\"u}tt, Heiko H and Kipnis, Alexander D and Diedrichsen, J{\"o}rn and Kriegeskorte, Nikolaus},
  year = {2023},
  journal = {eLife},
  volume = {12},
  pages = {e82566}
}

@article{shiWisdomPolarizedCrowds2019,
  title = {The Wisdom of Polarized Crowds},
  author = {Shi, Feng and Teplitskiy, Misha and Duede, Eamon and Evans, James A.},
  year = {2019},
  journal = {Nature Human Behaviour},
  volume = {3},
  number = {4},
  pages = {329--336},
  doi = {10.1038/s41562-019-0541-6}
}

@book{simonModelsBoundedRationality1982,
  title = {Models of {{Bounded Rationality}}},
  author = {Simon, Herbert A},
  year = {1982},
  publisher = {MIT Press},
  address = {Cambridge, Mass}
}

@book{simonModelsMan1957,
  title = {Models of {{Man}}},
  shorttitle = {Models of Man},
  author = {Simon, Herbert A},
  year = {1957},
  publisher = {Wiley},
  address = {New York}
}

@inproceedings{smythInferringGroundTruth1994,
  title = {Inferring {{Ground Truth}} from {{Subjective Labelling}} of {{Venus Images}}},
  booktitle = {Advances in {{Neural Information Processing Systems}}},
  author = {Smyth, Padhraic and Fayyad, Usama and Burl, Michael and Perona, Pietro and Baldi, Pierre},
  year = {1994},
  volume = {7},
  pages = {1085--1092},
  publisher = {MIT Press}
}

@article{sollOverconfidenceIntervalEstimates2004,
  title = {Overconfidence in Interval Estimates},
  author = {Soll, Jack B. and Klayman, Joshua},
  year = {2004},
  journal = {Journal of Experimental Psychology: Learning, Memory, and Cognition},
  volume = {30},
  number = {2},
  pages = {299--314},
  doi = {10.1037/0278-7393.30.2.299}
}

@article{stewartInformationGerrymanderingUndemocratic2019,
  title = {Information Gerrymandering and Undemocratic Decisions},
  author = {Stewart, Alexander J. and Mosleh, Mohsen and Diakonova, Marina and Arechar, Antonio A. and Rand, David G. and Plotkin, Joshua B.},
  year = {2019},
  journal = {Nature},
  volume = {573},
  number = {7772},
  pages = {117--121},
  doi = {10.1038/s41586-019-1507-6}
}

@article{stroopJudgmentGroupBetter1932,
  title = {Is the Judgment of the Group Better than That of the Average Member of the Group?},
  author = {Stroop, J. R.},
  year = {1932},
  journal = {Journal of Experimental Psychology},
  volume = {15},
  number = {5},
  pages = {550--562}
}

@book{surowieckiWisdomCrowds2004,
  title = {The {{Wisdom}} of {{Crowds}}},
  author = {Surowiecki, James},
  year = {2004},
  publisher = {Doubleday},
  address = {New York}
}

@article{tenenbaumTheorybasedBayesianModels2006,
  title = {Theory-Based {{Bayesian}} Models of Inductive Learning and Reasoning},
  author = {Tenenbaum, Joshua B. and Griffiths, Thomas L. and Kemp, Charles},
  year = {2006},
  journal = {Trends in Cognitive Sciences},
  volume = {10},
  number = {7},
  pages = {309--318},
  doi = {10.1016/j.tics.2006.05.009}
}

@article{trickBayesianClassifierFusion2022,
  title = {Bayesian {{Classifier Fusion}} with an {{Explicit Model}} of {{Correlation}}},
  author = {Trick, Susanne and Rothkopf, Constantin A},
  year = {2022},
  journal = {Proceedings of the 25th International Conference on Artificial Intelligence and Statistics (AISTATS) 2022, Valencia, Spain.},
  volume = {151},
  pages = {2282--2310}
}

@article{tsaiEffectsAmountInformation2008,
  title = {Effects of Amount of Information on Judgment Accuracy and Confidence},
  author = {Tsai, Claire I. and Klayman, Joshua and Hastie, Reid},
  year = {2008},
  journal = {Organizational Behavior and Human Decision Processes},
  volume = {107},
  number = {2},
  pages = {97--105},
  doi = {10.1016/j.obhdp.2008.01.005}
}

@article{turnerForecastAggregationRecalibration2014,
  title = {Forecast Aggregation via Recalibration},
  author = {Turner, Brandon M. and Steyvers, Mark and Merkle, Edgar C. and Budescu, David V. and Wallsten, Thomas S.},
  year = {2014},
  journal = {Machine Learning},
  volume = {95},
  number = {3},
  pages = {261--289},
  doi = {10.1007/s10994-013-5401-4}
}

@article{turnerWisdomCrowdApproach2011,
  title = {A Wisdom of the Crowd Approach to Forecasting},
  author = {Turner, Brandon M and Steyvers, Mark},
  year = {2011},
  journal = {NIPS Workshop on Computational Social Science and the Wisdom of Crowds},
  pages = {1--5.}
}

@article{tverskyJudgmentUncertaintyHeuristics1974,
  title = {Judgment under Uncertainty: {{Heuristics}} and Biases},
  author = {Tversky, Amos and Kahneman, Daniel},
  year = {1974},
  journal = {Science},
  volume = {185},
  number = {4157},
  pages = {1124--1131},
  doi = {10.1126/science.185.4157.1124}
}

@techreport{ungarGoodJudgmentProject2012,
  title = {The {{Good Judgment Project}}: {{A Large Scale Test}} of {{Different Methods}} of {{Combining Expert Predictions}}},
  author = {Ungar, Lyle and Mellors, Barb and Satop{\"a}{\"a}, Ville and Baron, Jon and Tetlock, Phil and Ramos, Jaime and Swift, Sam},
  year = {2012},
  number = {FS-12-06}
}

@article{vosoughiSpreadTrueFalse2018,
  title = {The Spread of True and False News Online},
  author = {Vosoughi, Soroush and Roy, Deb and Aral, Sinan},
  year = {2018},
  journal = {Science},
  volume = {359},
  number = {6380},
  pages = {1146--1151},
  doi = {10.1126/science.aap9559}
}

@article{wallstenComparingCalibrationCoherence1993,
  title = {Comparing the Calibration and Coherence of Numerical and Verbal Probability Judgments},
  author = {Wallsten, Thomas S. and Budescu, David V. and Zwick, Rami},
  year = {1993},
  journal = {Management Science},
  volume = {39},
  number = {2},
  pages = {176--190}
}

@article{wallstenPreferencesReasonsCommunicating1993,
  title = {Preferences and Reasons for Communicating Probabilistic Information in Verbal or Numerical Terms},
  author = {Wallsten, Thomas S. and Budescu, David V. and Zwick, Rami and Kemp, Steven M.},
  year = {1993},
  journal = {Bulletin of the Psychonomic Society},
  volume = {31},
  number = {2},
  pages = {135--138},
  doi = {10.3758/BF03334162}
}

@article{wallstenStateArtEncoding1983,
  title = {State of the Art---{{Encoding}} Subjective Probabilities: {{A}} Psychological and Psychometric Review},
  shorttitle = {State of the {{Art}}---{{Encoding Subjective Probabilities}}},
  author = {Wallsten, Thomas S. and Budescu, David V.},
  year = {1983},
  journal = {Management Science},
  volume = {29},
  number = {2},
  pages = {151--173},
  doi = {10.1287/mnsc.29.2.151}
}

@article{wheatleyEmergingScienceInteracting2023,
  title = {The {{Emerging Science}} of {{Interacting Minds}}},
  author = {Wheatley, Thalia and Thornton, Mark A. and Stolk, Arjen and Chang, Luke J.},
  year = {2023},
  journal = {Perspectives on Psychological Science},
  volume = {19},
  number = {2},
  pages = {355--373}
}

@article{winklerConsensusSubjectiveProbability1968,
  title = {The {{Consensus}} of {{Subjective Probability Distributions}}},
  author = {Winkler, Robert L.},
  year = {1968},
  journal = {Management Science},
  volume = {15},
  number = {2},
  pages = {B-61-B-75}
}

@article{zhengTruthInferenceCrowdsourcing2017,
  title = {Truth {{Inference}} in {{Crowdsourcing}}: {{Is}} the {{Problem Solved}}?},
  author = {Zheng, Yudian and Li, Guoliang and Li, Yuanbing and Shan, Caihua and Cheng, Reynold},
  year = {2017},
  journal = {Proceedings of the VLDB Endowment},
  volume = {10},
  number = {5},
  pages = {541--552}
}

\newpage
\section*{Supplementary information}

\subsection*{Idiosyncrasies of random ratings, majority vote, and median rating}
\textbf{Median beats majority vote for even numbers of ratings.} For odd numbers of ratings, the median rating is above 0.5 if and only if most ratings are above 0.5. The binary decisions rendered by the median rating and the majority vote will therefore be identical and their accuracy will match exactly. For even numbers of ratings, however, it can happen that the vote is equally split with half the ratings above and half below 0.5. In our definition of the majority vote, a random tie break is then performed. The median rating, by contrast, will average the two central ratings straddling 0.5. The median rating will fall on the side of 0.5 of the more confident one (the one farther from 0.5) of the two central ratings. It can, thus, take advantage of the continuous confidence information in the central two ratings in this rare scenario. The median rating therefore has slightly greater accuracy than the majority vote for even numbers of ratings, whereas it matches the majority vote in accuracy for odd numbers of ratings (Fig. \ref{fig:fig2}).

\textbf{For two ratings, the majority vote has the same accuracy as a random peer's rating.} The majority vote cannot benefit from the wisdom of a crowd of two people. It yields the same accuracy as picking a peer at random and trusting that single rating (left panel in Fig. \ref{fig:fig2}). In case both ratings fall on the same side of 0.5, majority vote and a random one of the two ratings lead to the same decision, so the expected accuracy is the same for both methods in this scenario. The alternative scenario, where one of the two ratings is above and the other below 0.5, entails a coin flip in majority vote. Choosing a random rating and using a coin flip to break the tie in the majority vote both yield chance performance. The overall accuracy of the two methods is therefore identical.




\subsection*{Independent Opinion Pool}
Given a set of  probability ratings $\mathbf{r}=(r_1,...,r_N)$ and assuming independent rating distributions conditional on the claim truth value as well as a flat truth prior, we have
\begin{align}
    \prod_{j=1}^N p(t|r_j)&=\prod_{j=1}^N \frac{p(r_j,t)}{p(r_j)}\\
    &=\prod_{j=1}^N\frac{p(r_j|t)p(t)}{p(r_j)}\\
    &=p(\mathbf{r}|t)p(t)\frac{p(t)^{N-1}}{\prod_{j=1}^N p(r_j)}\\
    &\propto p(t|\mathbf{r})
\end{align}
If we assume the raters are well-calibrated, i.e., $p(t=1|r_j)=r_j,$ then the lhs can be simplified to $\prod_{j=1}^N r_j$ for $t=1$ and  $\prod_{j=1}^N (1-r_j)$ for $t=0$.



\begin{figure}
    \centering
    \includegraphics[width=\textwidth]{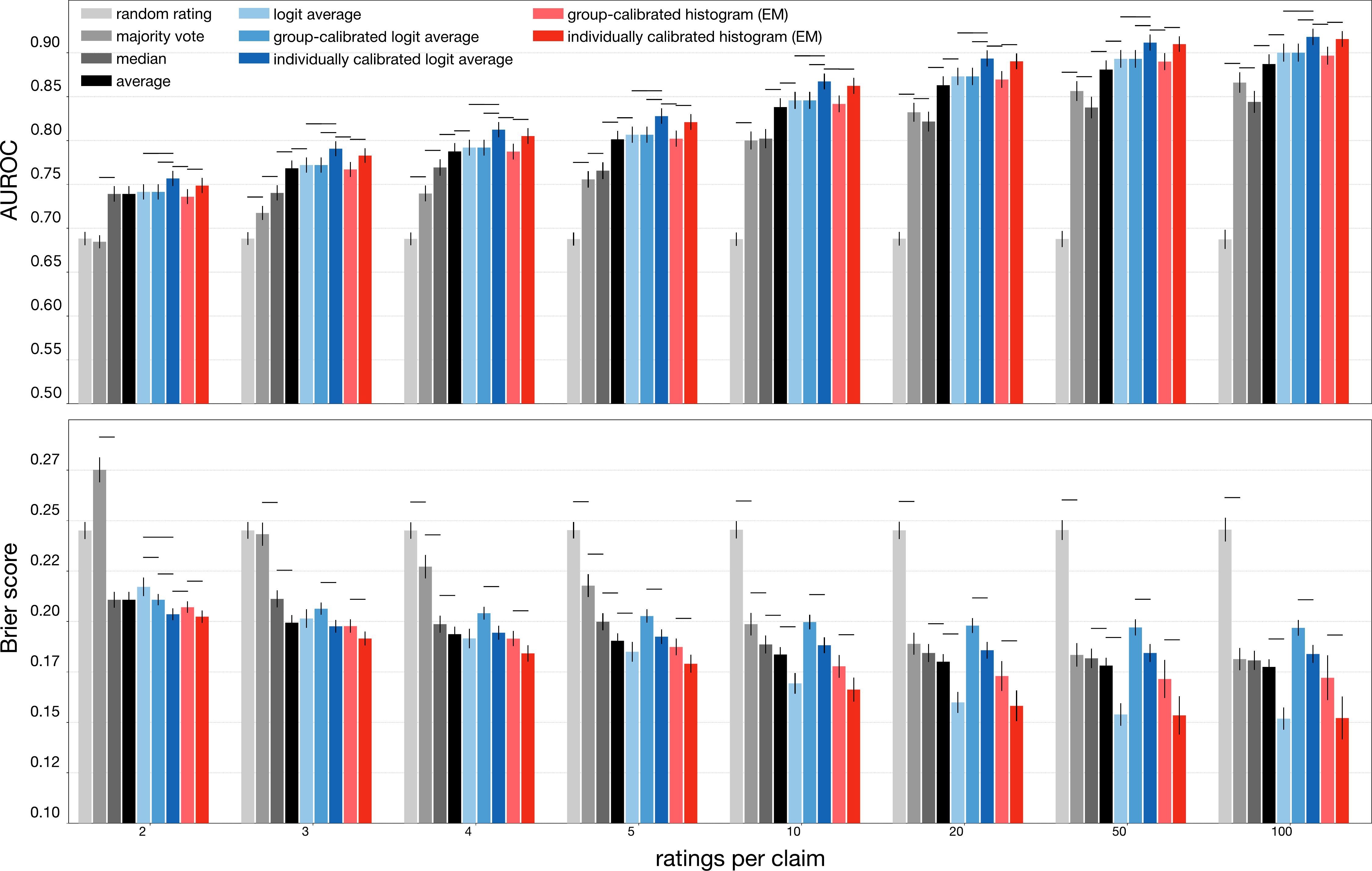}
    \caption{\footnotesize \textbf{Test auROC and Brier scores for collective-inference algorithms presented in Fig. 2.} Test auROC and Brier score of collective-inference algorithms for different numbers of ratings per claim. For neighboring bars, horizontal lines show significant differences (2-factor bootstrap, generalizing across both peers and claims, $p<0.05$, Bonferroni-corrected for 8 comparisons across numbers of ratings per claim, one-sided test for each pair of adjacent models of the hypothesis the more sophisticated model is better).}
\end{figure}


\begin{figure}
    \centering
    \includegraphics[width=\textwidth]{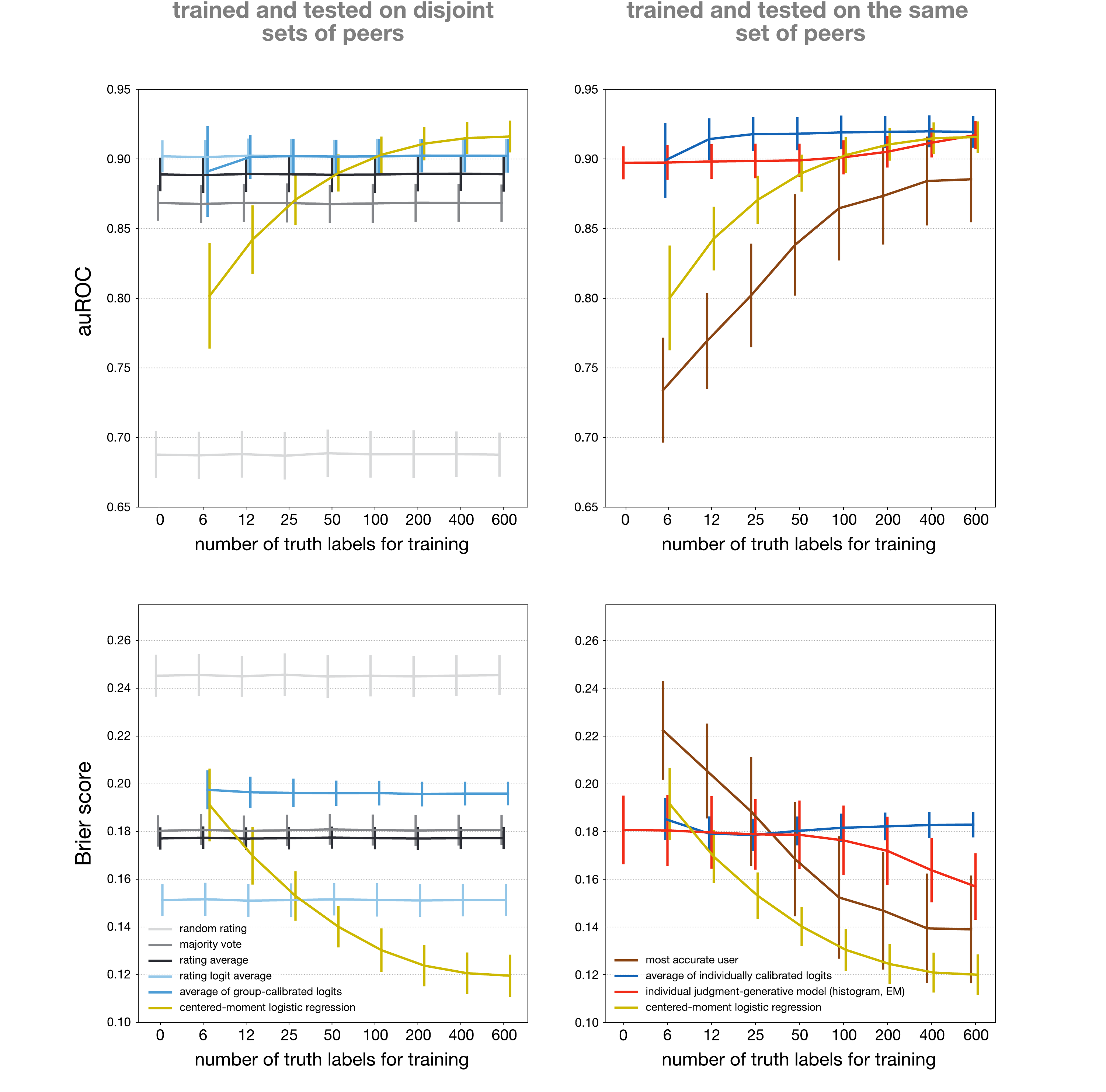}
    \caption{\footnotesize \textbf{Test auROC and Brier scores for collective-inference algorithms presented in Fig. 3.} Test auROC and Brier score of trained inference models under the scenario in which each user has rated each claim. The training and test sets each consist of 188 users (the same users across the split for the right panel and disjoint users for the left panel) and 600 disjoint claims. The number of truth-labeled claims randomly chosen from the training set is shown on the x-axis. The left panel shows the performance of models that do not model peers at the individual peer level, and the right panel shows the performance of models using probability ratings calibrated at the individual level. For the logistic regression model on the left panel, a disjoint set of peers is used for the training data to see how well the logistic regression method generalizes across disjoint sets of peers. Plots are shifted slightly on the x-axis to show error bars. Error bars represent standard error of the mean under a two-factor bootstrap resampling.}
\end{figure}

\begin{figure}
    \centering
    \includegraphics[width=\textwidth]{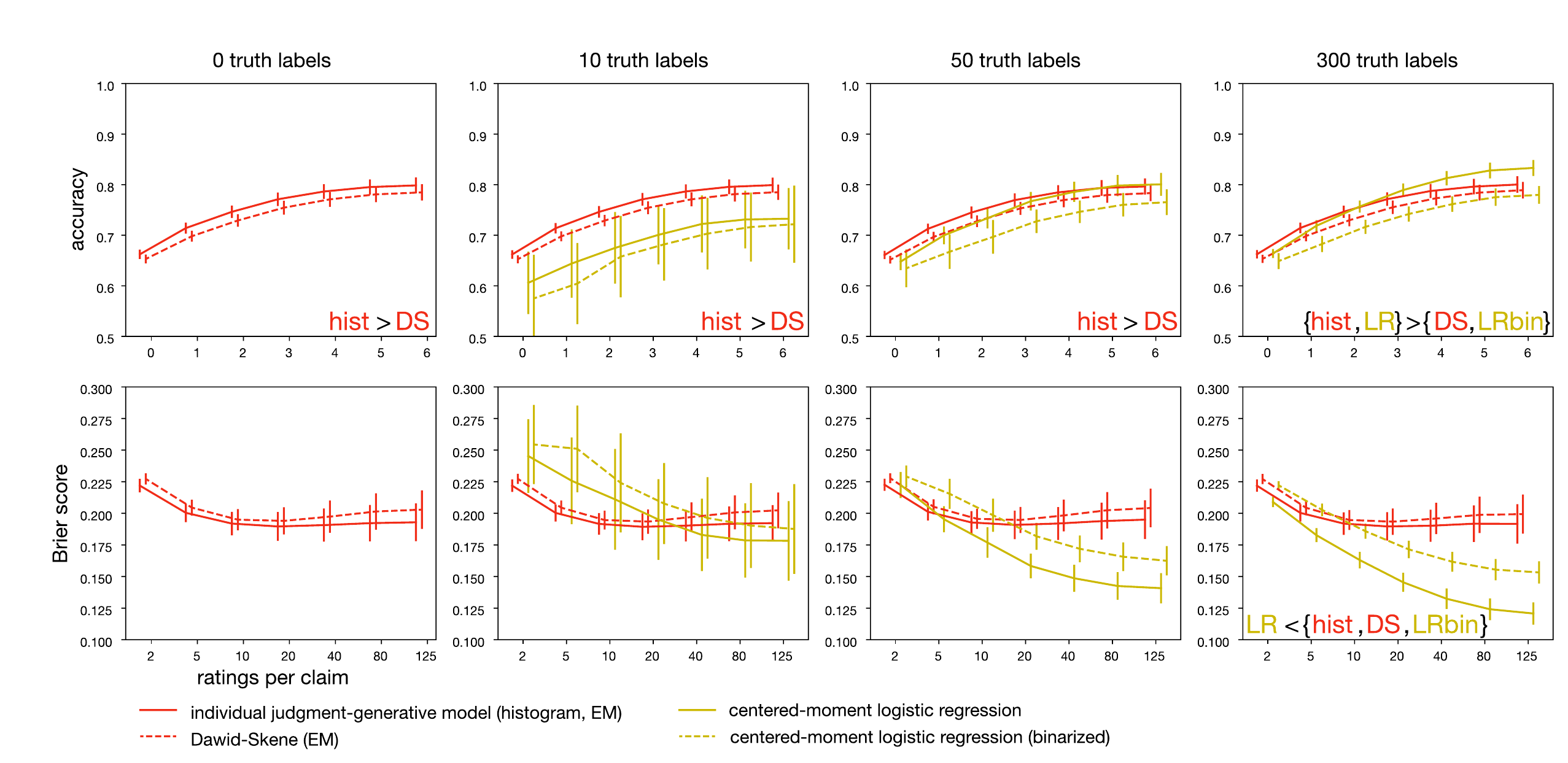}
    \caption{\footnotesize \textbf{Test auROC and Brier scores for collective-inference algorithms presented in Fig. 2.} Accuracy and Brier scores of selected truth inference models under thinning, in which a full probability rating matrix is made sparse by randomly splitting each row and/or column into multiple sparse rows or columns, respectively. Models are learned using a dataset of 188 peers and 600 claims. The number claims provided with ground truth labels differs across panels. Performance is reported from model predictions of the unlabeled portion of the dataset.}
    \label{fig:continuousVsBinarized_accBrier}
\end{figure}

\begin{figure}
    \centering
    \includegraphics[width=1.0\textwidth]{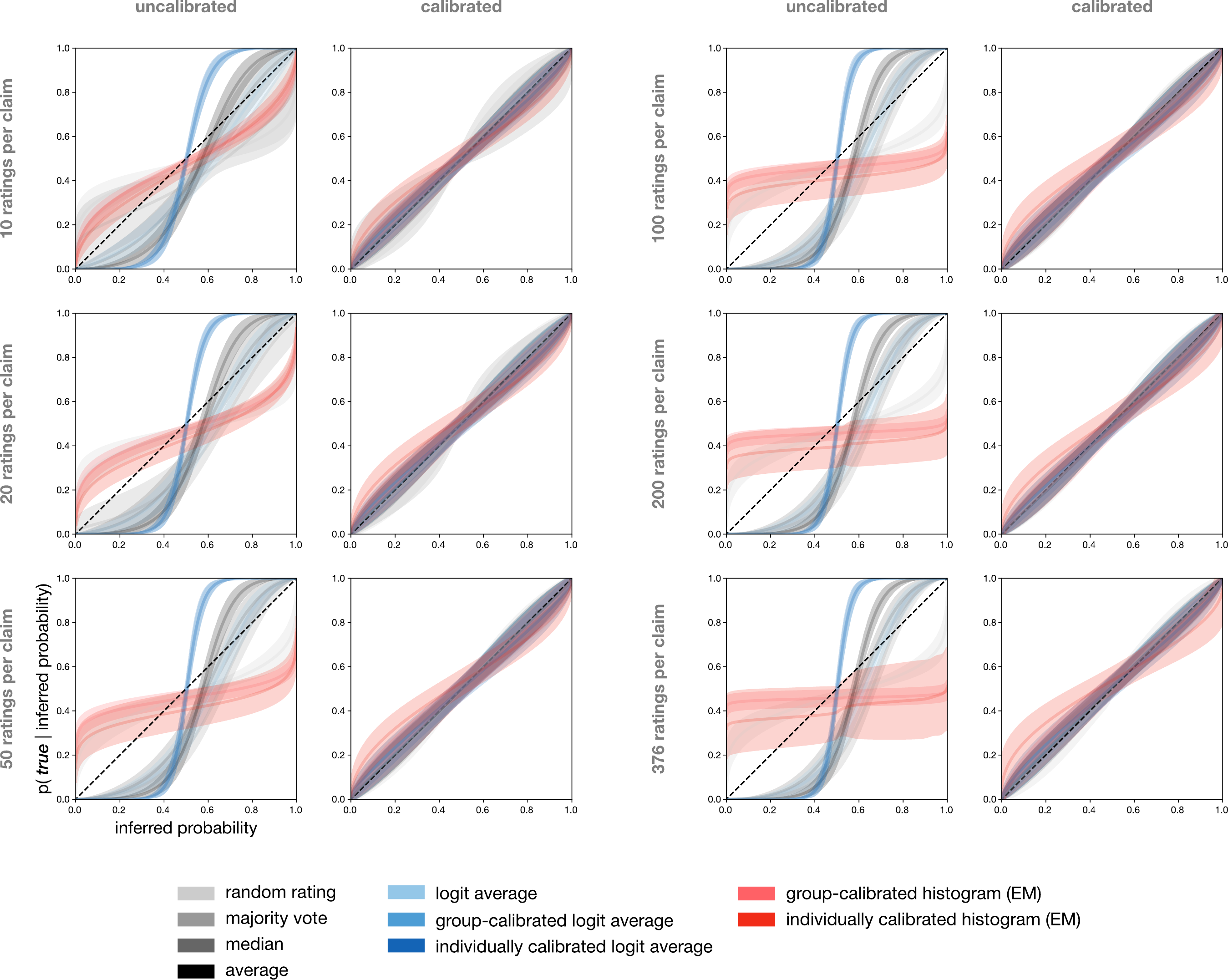}
    \caption{\footnotesize \textbf{Experiment in Fig. \ref{fig:fig5} for a wider range of number of peers}. Inference algorithms are trained on a random set consisting of 100 labeled claims and 50 unlabeled claims. A logistic model is fit to the resulting inference estimates on a test set of 950 claims and plotted to estimate the calibration function of the inference algorithms (thick lines). Lines overlapping with the identity line are better calibrated. We then train a logistic regression model on a labeled calibration set of each inference algorithm's estimates of 100 labeled claims to learn a mapping to better calibrate each algorithm's estimate. The procedure is performed on a random split 100 times with thick lines representing mean $+/-$ standard deviation.}
    \label{fig:fig12}
\end{figure}

\begin{figure}
    \centering
    \includegraphics[width=1.0\textwidth]
    {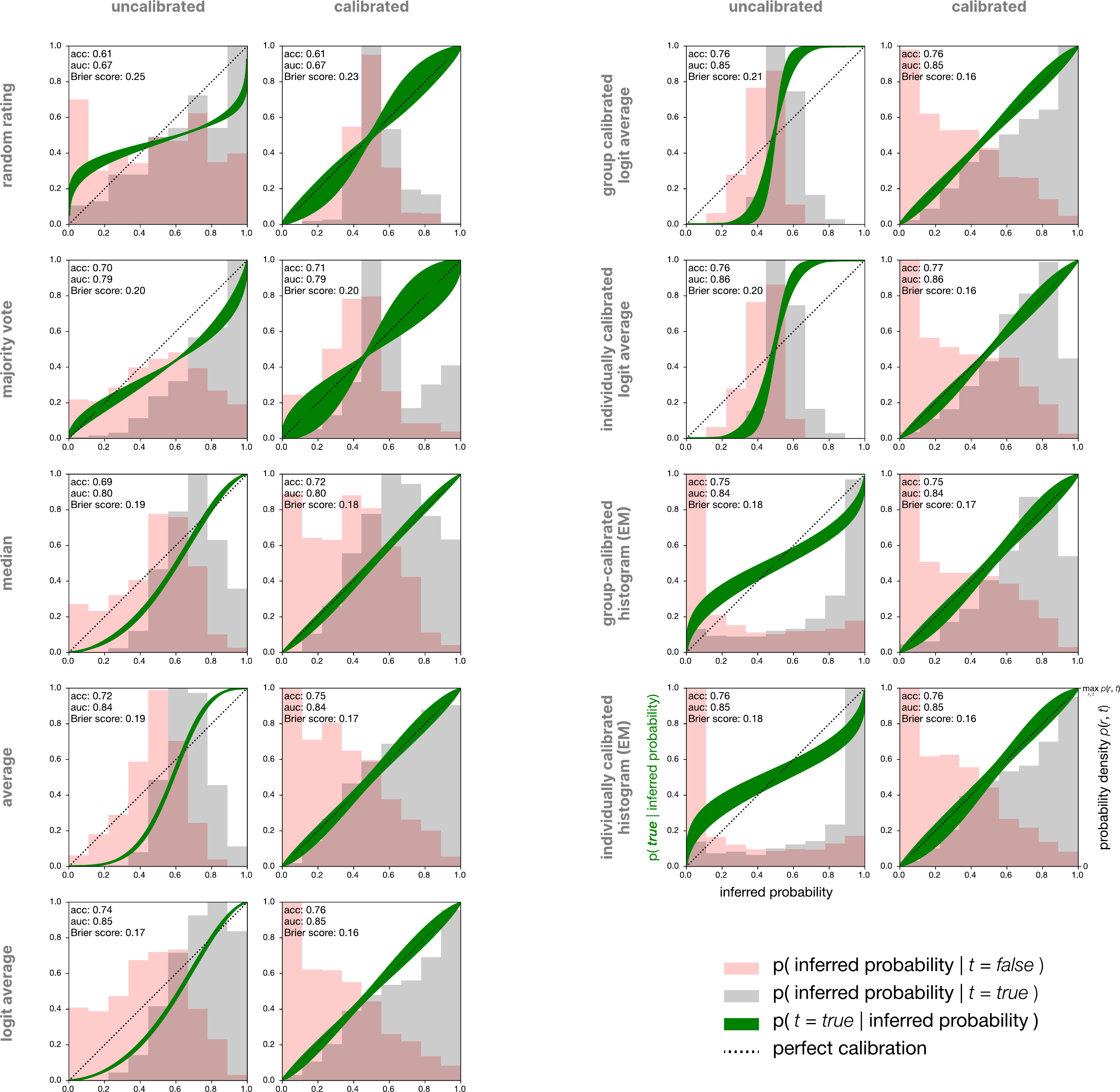}
    \caption{\footnotesize \textbf{Separate calibration plots and histograms in Fig. \ref{fig:fig12}.} Calibration plots for collective inference algorithms for 10 ratings per claim in Fig. \ref{fig:fig12} with each inference algorithm's empirical predictive distribution for true (gray) and false (pink) claims. Calibration line width represents standard deviation over 100 random samplings of peers and random partitionings of the unlabeled training (100 points), labeled training (50 points), calibration (100 points), and test (950) sets.}
    \label{fig:fig13}
\end{figure}

\begin{figure}[h]
\centering
\includegraphics[width=1.0\textwidth]
    {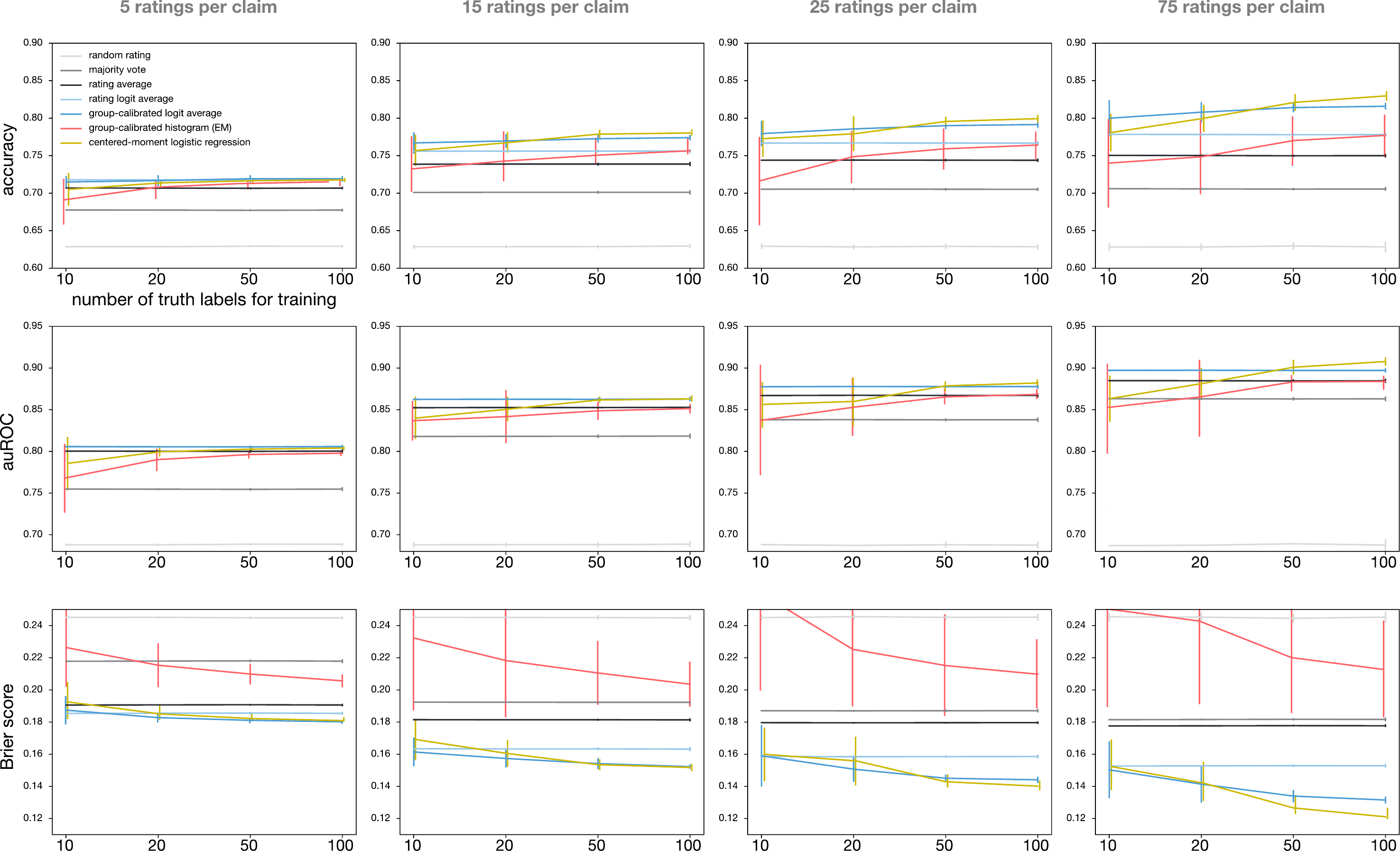}
\caption{\footnotesize \textbf{Performance of collective inference algorithms without peer-specific learning (dataset thinned to simulate one rating per peer).} Error bars represent standard error over 100 draws of a 2-factor bootstrap over peers and claims and a random partitioning of the data into a labeled training set with varying size (``number of fixed truths" on the x-axis) and a test set consisting of the rest of the claims.}
\label{fig:fig14}
\end{figure}
\end{document}